\documentclass[12pt]{article}

\pdfoutput=1
\usepackage[latin9]{inputenc}
\usepackage[top=80pt,bottom=85pt,left=67pt,right=67pt]{geometry}
\usepackage{amssymb,amsmath}
\usepackage{xypic,subfigure}
\usepackage{graphicx,color}
\usepackage{float}
\usepackage{cite}
\usepackage[debug,pageanchor=false]{hyperref}
\definecolor{link}{rgb}{.8,.15,.1}
\hypersetup{colorlinks=true,linkcolor=link,citecolor=link,urlcolor=link,linktocpage}

\makeatletter
\@addtoreset{equation}{section}
\makeatother

\renewcommand{\theequation}{\thesection.\arabic{equation}}

\newcommand{\beq}{\begin{equation}}
\newcommand{\eeq}{\end{equation}}
\newcommand{\bea}{\begin{eqnarray}}
\newcommand{\eea}{\end{eqnarray}}

\newcommand{\eq}{\begin{equation}}
\newcommand{\feq}{\end{equation}}
\newcommand{\eqn}{\begin{eqnarray}}
\newcommand{\feqn}{\end{eqnarray}}

\newcommand{\ma}[1]{\mbox{$\mathcal{#1}$}}

\newcommand{\mrm}[1]{\mbox{$\mathrm{#1}$}}

\begin{document}
\begin{titlepage}

\begin{center}

\vskip .9in 
\noindent

{\Large \bf{New AdS$_2$ supergravity duals of 4d SCFTs with defects}}

\bigskip\medskip

Yolanda Lozano\footnote{ylozano@uniovi.es},  Nicol\`o Petri\footnote{petrinicolo@uniovi.es}, Cristian Risco\footnote{cristianrg96@gmail.com} \\

\bigskip\bigskip
{\small 

Department of Physics, University of Oviedo,
Avda. Federico Garcia Lorca s/n, \\ 33007 Oviedo, Spain}

\vskip .2cm

{\small

Instituto Universitario de Ciencias y Tecnolog\'ias Espaciales de Asturias (ICTEA),\\
Calle de la Independencia 13, 33004 Oviedo, Spain}

\vskip 2cm 

     	{\bf Abstract }
     	\end{center}
     	\noindent
	
	We construct new families of $\text{AdS}_2 \times S^2 \times S^2$ solutions with 4 supercharges in Type II supergravities. We show that subclasses of these solutions can be interpreted in terms of defect branes embedded in 4d $\mathcal{N} = 4 $ SYM, or orbifolds thereof. This is explicitly realised by showing that the solutions asymptote locally to $\text{AdS}_5 \times S^5/\mathbb{Z}_n$, in Type IIB, or its T-dual background, in Type IIA. The latter is a Gaiotto-Maldacena geometry realised on an intersection of D4 and NS5 branes. We extend the Type IIA solutions to include D6 branes, and interpret them as describing backreacted baryon vertices within 4d $\mathcal{N} = 2$ CFTs living in D4-NS5-D6 intersections. We propose explicit quiver quantum mechanics in which the defect branes play the role of colour branes, with the D4 branes of the D4-NS5-D6 intersection becoming flavour branes. These quivers are used to compute the degeneracies of the ground states of the dual super conformal quantum mechanics, that are shown to agree with the holographic expressions.	
	

\noindent

\vfill
\eject

\end{titlepage}

\setcounter{footnote}{0}

\tableofcontents

\setcounter{footnote}{0}
\renewcommand{\theequation}{{\rm\thesection.\arabic{equation}}}

\section{Introduction}

Ever since the first formulation of the AdS/CFT correspondence \cite{Maldacena:1997re}, the construction of AdS vacua and their field theory duals has become the most concrete approach to understand the physics of extended objects in string theory. Among the extremely varied research lines opened in holography in the last twenty years the study of AdS$_3$ and AdS$_2$ vacua and their dual realisations has deserved special attention. This is highly relevant given the role played by these geometries as horizons of black objects in (super)gravity \cite{Strominger:1996sh}. However, despite the enormous progress made on lower-dimensional AdS holography, a completely consistent and well-understood string theory description is still missing.

On the gravity side a great effort in classifying AdS$_3$ and AdS$_2$ vacua has been done in old and recent literature (see 
\cite{Kim:2005ez}-\cite{Lozano:2021rmk}
for a non-exhaustive list of references). The underlying brane description of these solutions, key to the study of their CFT interpretation, becomes more complicated as the dimensionality of the internal space increases, due to the richer structure of the possible geometries and fluxes.
In this paper we will focus our study on AdS$_2$ spaces. We will present new classes of $\ma N=4$ solutions with a clear underlying brane interpretation. Taking these fully-backreacted backgrounds as starting points, we will then turn to the study of their dual field theory interpretation.

It is well-known that the AdS$_2$/CFT$_1$ correspondence has important technical and conceptual problems. The fact that the boundary of AdS$_2$ is not-connected poses the problem of where the dual super conformal quantum mechanics (SCQM) lives, and the associated non-factorisability of the quantum partition function (see for example \cite{Harlow:2018tqv}). In this paper we will consider AdS$_2$/CFT$_1$ as an effective lower dimensional description of a given brane intersection, described in the UV in higher dimensions. This was, for example, the approach followed in \cite{Dibitetto:2019nyz}.

A very powerful tool in this context is given by defect conformal field theories and their brane engineering (for a non-exhaustive list of references see \cite{Karch:2000gx}-\cite{Gutperle:2020rty}
These theories resolve the dynamical degrees of freedom of a holographic CFT associated with an intersection of \emph{defect branes} with a bound state of \emph{background branes}, in which a higher dimensional CFT lives.
Holographically this situation is well-described by AdS solutions with non-compact internal manifolds, reproducing a locally higher-dimensional AdS geometry asymptotically.  In such cases the divergence of the  holographic free energy (or central charge) is interpreted as the need of a UV completion into higher-dimensions, rather than as a pathology of the theory. From the gravity side this divergent behaviour is resolved within the geometry of the higher-dimensional AdS vacuum, while from the gauge theory side the defect CFT is realised in terms of a position-dependent coupling within a higher-dimensional CFT whose conformal isometries (and supersymmetries) are explicitly broken.

We start our investigations in section \ref{IIApart}, where we consider a F1-D2-D4$'$-NS5$'$ brane intersection in Type IIA string theory \cite{Boonstra:1997dy,Boonstra:1998yu} ending on a bound state of D4-NS5 branes. 
We construct the general brane solution associated to this brane system, from where we extract the near horizon geometry, consisting on a fully-backreacted $\mrm{AdS}_2\times S^2$ spacetime. In this way we derive a new class of $\ma N=4$ $\mrm{AdS}_2\times S^2\times S^2\times \mathbb R^2\times S^1$ geometries foliated over a line in Type IIA supergravity. The main property of this class of solutions is that they are completely determined by the dynamics of the D4-NS5 branes wrapping the curved AdS$_2\times S^2$ geometry.
We then concentrate on the situation in which the D4-NS5 branes are described by the semi-localised solution of \cite{Youm:1999ti}-\cite{Oz:1999qd}. We show that in  this case the AdS$_2$ solution is resolved in the UV within the AdS$_5$ geometry arising in the near horizon of the semi-localised D4-NS5 branes.  This is a Gaiotto-Maldacena geometry related to the Type IIB $\text{AdS}_5\times S^5/\mathbb{Z}_n$ background upon T-duality, and is therefore holographically dual to a $\mathbb{Z}_n$ orbifold of 4d $\ma N=4$ SYM. Based on this we propose an interpretation to the AdS$_2$ solution as describing backreacted line defects within a $\mathbb{Z}_n$ orbifold of 4d $\ma N=4$ SYM.  

In section \ref{IIBpart} we T-dualise the brane set-up studied in Type IIA along the circular direction on which the semi-localised D4 branes are stretched. The brane picture in Type IIB becomes an 
F1-D1-NS5-D5 brane intersection ending on D3 branes probing an A-type singularity. Close to the horizon a new class of $\ma N=4$ $\mrm{AdS}_2\times S^2\times S^2\times \mathbb R^2\times S^1$ geometries foliated over a line arises. These solutions describe the near horizon regime of D3 branes intersecting with KK monopoles, wrapping the fully-backreacted geometry associated to the F1-D1-NS5-D5 branes. We show that the defect interpretation found in Type IIA  is maintained, with the $\mrm{AdS}_2$ geometry now resolved within the $\mrm{AdS}_5\times S^5/\mathbb{Z}_n$ vacuum associated to the D3 branes on the A-type singularity. The case $n=1$, corresponding to the absence of KK monopoles, arises when the AdS$_2$ solution is resolved in the UV within the  $\mrm{AdS}_5\times S^5$ Freund-Rubin solution. In this case the AdS$_2$ solution finds an interpretation as describing backreacted line defects within 4d $\ma N=4$ SYM. 

In section \ref{IIAwithD6} we include D6 branes in the Type IIA brane set-up studied in section \ref{IIApart}. By doing this we extend the class of $\ma N=4$ $\mrm{AdS}_2\times S^2\times S^2\times \mathbb R^2\times S^1$ geometries constructed therein to describe the near horizon of D4-NS5-D6 branes wrapping the fully-backreacted geometry associated to the F1-D2-D4$'$-NS5$'$ branes. Even if in this case we have not been able to construct explicit solutions that asymptote to the Gaiotto-Maldacena geometry associated to the D4-NS5-D6 intersection we perform a detailed study of the 1d CFT dual to the solutions, and explicitly propose quiver quantum mechanics that flow to SCQM in the IR. We compute the ``central charge'' of these SCQMs, that we interpret as the degeneracy of their ground states, and show that it agrees with the holographic result. We propose an interpretation for the quiver quantum mechanics as describing D4$'$-D2 baryon vertices within the 4d $\ma N=2$ CFT living in the D4-NS5-D6 brane intersection.

Finally, section \ref{conclusions} contains our conclusions and future directions. Appendix \ref{appMtheory} includes the strong coupling, M-theory realisation of the brane intersection and near horizon geometries discussed in section \ref{IIApart}.

\section{AdS$_2$ solutions in Type IIA  as defects within AdS$_5$}\label{IIApart}

In this section we present a new class of AdS$_2$ solutions to Type IIA string theory that arise in the near horizon limit of a brane set-up consisting of F1-D2-D4$'$-NS5$'$ branes ending on a D4-NS5 bound state. These brane configurations are described close to the horizon by $\ma N=4$ $\mrm{AdS}_2\times S^2\times S^2\times \mathbb R^2\times S^1$ geometries foliated over a line. We show that a suitable prescription for the distributions of charges of the D4-NS5 branes produces a solution within this class that asymptotes locally to a AdS$_5$ vacuum in Type IIA, associated to the D4-NS5 brane intersection. This allows one to interpret this solution as describing a line defect CFT within the $\ma N=2$ 4d CFT dual to the aforementioned AdS$_5$ vacuum. 

\subsection{F1-D2-D4$'$-NS5$'$-D4-NS5 intersecting branes} \label{branesystemIIA}

Our starting point is the brane set-up depicted in Table \ref{Table:branesinmasslessIIA1}. This is a BPS/8 brane intersection that can be interpreted in terms of a F1-D2-D4'-NS5' brane intersection ending on a BPS/4 bound state of D4-NS5 branes.
\begin{table}[http!]
\renewcommand{\arraystretch}{1}
\begin{center}
\scalebox{1}[1]{
\begin{tabular}{c c cc c|| c  c  c| c c c}
 branes & $t$ & $\rho$ & $\varphi^1$ & $\varphi^{2}$ & $y$ & $z$ & $\psi$ & $r$ & $\theta^1$ & $\theta^2$ \\
\hline \hline
$\mrm{D}4$ & $\times$ & $\times$ & $\times$ & $\times$ & $-$ & $-$ & $\times$ & $-$ & $-$ & $-$ \\
$\mrm{NS}5$ & $\times$ & $\times$ & $\times$ & $\times$ & $\times$ & $\times$ & $-$ & $-$ & $-$ & $-$ \\
$\mrm{F}1$ & $\times$ & $-$ & $-$ & $-$ & $-$ & $\times$ & $-$ & $-$ & $-$ & $-$ \\
$\mrm{D}2$ & $\times$ & $-$ & $-$ & $-$ & $\times$ & $-$ & $\times$ & $-$ & $-$ & $-$ \\
$\mrm{D}4'$ & $\times$ & $-$ & $-$ & $-$ & $\times$ & $-$ & $-$ & $\times$ & $\times$ & $\times$ \\
$\mrm{NS}5'$ & $\times$ & $-$ & $-$ & $-$ & $-$ & $\times$ & $\times$ & $\times$ & $\times$ & $\times$ \\
\end{tabular}
}
\caption{BPS/8 intersection describing F1-D2-D4$'$-NS5$'$ branes ending on a D4-NS5 bound state. This system defines a $\ma N=4$ line defect SCQM within the $\ma N=2$ 4d CFT living in the D4-NS5 branes.} \label{Table:branesinmasslessIIA1}
\end{center}
\end{table}

We take the F1-D2-D4$'$-NS5$'$ branes completely localised within the four dimensional worldvolume of the orthogonal D4-NS5 branes. As shown in \cite{Faedo:2020nol,Faedo:2020lyw}, this requirement is crucial\footnote{At least for AdS$_2$ and AdS$_3$ vacua.} in order to decouple the field equations of the F1-D2-D4$'$-NS5$'$  {\it defect} branes from those of the D4-NS5 {\it background} branes. Besides this we take the  
D4 branes completely localised in their transverse space and stretched within the NS5-branes in a circular direction, along which the NS5-branes are smeared.

The metric and dilaton for such a system enjoy the following form,
\begin{equation}
\label{brane_metric_D4NS5D2F1D4}
\begin{split}
d s_{10}^2 &= H_{\mathrm{D}4}^{-1/2}  \left[-H_{\mathrm{F}1}^{-1}H_{\mathrm{D}2}^{-1/2} H_{\mathrm{D}4'}^{-1/2} \,dt^2+H_{\mathrm{D}2}^{1/2} H_{\mathrm{D}4'}^{1/2}H_{\mathrm{NS}5'} \bigl(d\rho^2+\rho^2ds^2_{S^2}\bigr) \right] \\
&+ H_{\mathrm{D}4}^{1/2} \left[H_{\mathrm{D}2}^{-1/2} H_{\mathrm{D}4'}^{-1/2}H_{\mathrm{NS}5'}\,dy^2+H_{\mathrm{F}1}^{-1} H_{\mathrm{D}2}^{1/2} H_{\mathrm{D}4'}^{1/2}\,dz^2 \right]\\
&+H_{\mathrm{NS}5}H_{\mathrm{D}4}^{-1/2}H_{\mathrm{D}2}^{-1/2} H_{\mathrm{D}4'}^{1/2}\,d\psi^2+H_{\mathrm{NS}5}H_{\mathrm{D}4}^{1/2}H_{\mathrm{D}2}^{1/2} H_{\mathrm{D}4'}^{-1/2}\bigl(dr^2+r^2ds^2_{\tilde S^2}\bigr)\,,\\
e^{\Phi}&= H_{\mathrm{NS}5}^{1/2}H_{\mathrm{D}4}^{-1/4}H_{\mathrm{F}1}^{-1/2}H_{\mathrm{D}2}^{1/4}H_{\mathrm{D}4'}^{-1/4}H_{\mathrm{NS}5'}^{1/2}\,.
\end{split}
\end{equation}
Here $S^2$ is the 2-sphere spanned by the coordinates $(\varphi^1,\varphi^2)$ in Table \ref{Table:branesinmasslessIIA1}, and $\tilde S^2$ the 2-sphere spanned by $(\theta^1,\theta^2)$.
The condition that the F1-D2-D4$'$-NS5$'$ branes are completely localised within the worldvolume of the orthogonal D4-NS5 branes implies that $H_{\mathrm{F}1}$, $H_{\mathrm{D}2}$, $H_{\mathrm{D}4'}$ and $H_{\mathrm{NS}5'}$ depend only on  $\rho$. Besides, as we mentioned above, we ask that $H_{\mathrm{D}4}=H_{\mathrm{D}4}(y,z,r)$ and $H_{\mathrm{NS}5}=H_{\mathrm{NS}5}(r)$. Namely, we take the D4 branes completely localised in their transverse space and the NS5 branes smeared along the $\psi$ direction. With these prescriptions the fluxes take the form
\begin{equation}
\begin{split}\label{fluxes_D4NS5D2F1D4}
H_{(3)} &= -\partial_\rho H_{\mathrm{F}1}^{-1}dt\wedge d\rho\wedge dz+ \partial_\rho H_{\mathrm{NS}5'}\,\rho^2\, \text{vol}_{  S^2}\wedge dy+ \partial_r H_{\mathrm{NS}5}\,r^2\,d\psi \wedge\text{vol}_{ \tilde S^2}\,,\\
F_{(4)}&=\partial_\rho H_{\mathrm{D}2}^{-1}dt\wedge d\rho\wedge dy \wedge d\psi+ \partial_\rho H_{\mathrm{D}4'}\,\rho^2\, \text{vol}_{  S^2}\wedge dz\wedge d\psi+\partial_r H_{\mathrm{D}4}\,r^2\,  dy\wedge dz \wedge \text{vol}_{ \tilde S^2}\\
&+H_{\mathrm{D}2}H_{\mathrm{NS}5'}^{-1}H_{\mathrm{NS}5}\partial_y H_{\mathrm{D}4}\,r^2\,  dz\wedge d r \wedge \text{vol}_{ \tilde S^2}
-H_{\mathrm{F}1}H_{\mathrm{D}4'}^{-1}H_{\mathrm{NS}5}\partial_z H_{\mathrm{D}4}\,r^2\,  dy\wedge d r \wedge \text{vol}_{ \tilde S^2}\,.\\
\end{split}
\end{equation}
It can be seen that the equations of motion and Bianchi identities for \eqref{brane_metric_D4NS5D2F1D4} and \eqref{fluxes_D4NS5D2F1D4} decouple into two groups. The equations for F1-D2-D4$'$-NS5$'$ branes are equivalent to
\begin{equation} \label{D4NS5D2F1EOM}
\begin{split}
&\nabla^2_{\mathbb{R}^3_\rho} H_{\mathrm{D}2}=0 \qquad \text{with}\qquad  H_{\mathrm{NS}5'}=H_{\mathrm{D}2}\,,\\
&\nabla^2_{\mathbb{R}^3_\rho} H_{\mathrm{F}1}=0 \qquad \text{with}\qquad  H_{\mathrm{D}4'}=H_{\mathrm{F}1}\,,\\
\end{split}
\end{equation}
while the D4-NS5 system satisfies the equations
\begin{equation}\label{EOM_NS5D4}
 \nabla^2_{\mathbb{R}^3_r} H_{\mathrm{D}4}+H_{\mathrm{NS}5}\nabla^2_{\mathbb{R}^2_{(y,z)}} H_{\mathrm{D}4}=0\qquad \text{and} \qquad  \nabla^2_{\mathbb{R}^3_r} H_{\mathrm{NS}5}=0\,.
\end{equation}
Note that the conditions \eqref{D4NS5D2F1EOM} do not impose any warping within the 2d subspace $\mathbb{R}^2_{(y,z)}$ parametrised by $y$ and $z$. This happens because the defect branes are completely smeared within this subspace.
 

In order to extract the AdS$_2$ geometry we choose the particular solutions
\begin{equation}\label{stringsol}
  H_{\mathrm{D}2}=1+\frac{q_{\mathrm{D}2}}{\rho} \qquad \text{and} \qquad H_{\mathrm{F}1}=1+\frac{q_{\mathrm{F}1}}{\rho}\,,
\end{equation}
where $q_{\mathrm{D}2}$ and $q_{\mathrm{F}1}$ are integration constants related to the quantised charges of the defect branes. Taking the limit $\rho \rightarrow 0$ we obtain\footnote{In order to reproduce unitary AdS$_2$ at the horizon we rescaled the time as $t\rightarrow q_{\mathrm{D}2} q_{\mathrm{F}1}t$.},
\begin{equation}
\label{brane_metric_D4NS5D2F1D4_nh}
\begin{aligned}
ds_{10}^2 &= q_{\mathrm{D}2}^{3/2}q_{\mathrm{F}1}^{1/2}    H_{\mathrm{D}4}^{-1/2}\left(ds^2_{\text{AdS}_2}+ds^2_{S^2}\right) +q_{\mathrm{D}2}^{1/2}q_{\mathrm{F}1}^{-1/2} H_{\mathrm{D}4}^{1/2}\,  \left( dy^2+dz^2 \right)\\
&+q_{\mathrm{D}2}^{-1/2} q_{\mathrm{F}1}^{1/2}H_{\mathrm{NS}5}H_{\mathrm{D}4}^{-1/2}\,d\psi^2+H_{\mathrm{NS}5}H_{\mathrm{D}4}^{1/2}q_{\mathrm{D}2}^{1/2} q_{\mathrm{F}1}^{-1/2}\bigl(dr^2+r^2ds^2_{\tilde S^2}\bigr)\,,\\
 e^{\Phi}&=q_{\mathrm{D}2}^{3/4} q_{\mathrm{F}1}^{-3/4}\,H_{\mathrm{NS}5}^{1/2}H_{\mathrm{D}4}^{-1/4}\,,\\
  H_{(3)} &= -q_{\mathrm{D}2}\text{vol}_{\text{AdS}_2}\wedge dz-q_{\mathrm{D}2} \text{vol}_{  S^2}\wedge dy+ \partial_r H_{\mathrm{NS}5}\,r^2\,d\psi \wedge\text{vol}_{ \tilde S^2}\,,\\
F_{(4)}&=q_{\mathrm{F}1}\text{vol}_{\text{AdS}_2}\wedge dy \wedge d\psi-q_{\mathrm{F}1}\, \text{vol}_{  S^2}\wedge dz\wedge d\psi+\partial_r H_{\mathrm{D}4}\,r^2\,  dy\wedge dz \wedge \text{vol}_{ \tilde S^2}\\
&+H_{\mathrm{NS}5}\partial_y H_{\mathrm{D}4}\,r^2\,  dz\wedge d r \wedge \text{vol}_{ \tilde S^2}
-H_{\mathrm{NS}5}\partial_z H_{\mathrm{D}4}\,r^2\,  dy\wedge d r \wedge \text{vol}_{ \tilde S^2}\,.\\
\end{aligned}
\end{equation}
 These geometries represent a new class of backgrounds of the form $\mrm{AdS}_2\times S^2\times {\tilde S}^2\times \mathbb R^2_{(y,z)}\times S^1_\psi$, foliated over an interval, parametrised by $r$. 
The functions $H_{\mathrm{D}4}(y,z,r)$ and $H_{\mathrm{NS}5}(r)$ are solutions to the equations \eqref{EOM_NS5D4}, and describe a D4-NS5 bound state localised in the subspace $\mathbb{R}^2_{(y,z)}\times \mathbb{R}^3_r$, with $\mathbb{R}^3_r$ spanned by $r$ and ${\tilde S}^2$. The backgrounds \eqref{brane_metric_D4NS5D2F1D4_nh} constitute a vast class of $\ma N=4$ solutions to Type IIA string theory, determined by the charge distribution of the D4-NS5 system. In the next section we will analyse a remarkable example in which a particular solution for $H_{\mathrm{D}4}$ and $H_{\mathrm{NS}5}$ gives rise to a $\mrm{AdS}_2\times S^2$ background reproducing asymptotically locally an $\mrm{AdS}_5$ vacuum related by T-duality to $\mrm{AdS}_5\times S^5/\mathbb{Z}_n$.

\subsection{D4-NS5 branes and warped AdS$_5$}\label{AdS5semilocalD4NS5}

As we have just mentioned, a key property of the brane system depicted in Table \ref{Table:branesinmasslessIIA1} is the possibility of decoupling the dynamics of the F1-D2-D4$'$-NS5$'$ defect branes from that of the D4-NS5 system. This is manifest at the level of the equations of motion, with the equations in \eqref{D4NS5D2F1EOM} describing the F1-D2-D4$'$-NS5$'$ subsystem and those in \eqref{EOM_NS5D4} the D4-NS5 background branes.
The simplest situation in which one can exploit this property is when the F1-D2-D4$'$-NS5$'$ defect branes of Table \ref{Table:branesinmasslessIIA1} are ``zoomed out'', namely, just the D4-NS5 bound state is considered. This is done  at the level of the brane solution \eqref{brane_metric_D4NS5D2F1D4} by taking the limit $\rho \rightarrow +\infty$ in \eqref{stringsol}, and looking at the resulting backreacted solution. In this subsection we will be interested in this limit. We will produce an explicit AdS$_5$ solution associated to the D4-NS5 subsystem, on which F1-D2-D4$'$-NS5$'$ branes will later be embedded to produce a backreacted AdS$_2$ solution in the class given by \eqref{brane_metric_D4NS5D2F1D4_nh}.

The D4-NS5 system has been extensively studied in the literature. Field theoretically it was first studied in \cite{Witten:1997sc}. The set-up contains NS5 branes extended in the directions $(\mathbb{R}^{1,3}, x_4,x_5)$ at different positions $x_{6,n}$ in the $x_6$-direction, and D4 branes extended in 
$(\mathbb{R}^{1,3}, x_6)$ in between the NS5 branes. This brane set-up preserves 1/4 of the supersymmetries.  The field theory living in the (colour) D4 branes is effectively four dimensional at low energies compared to the inverse of the separation between the NS5 branes, with the effective gauge coupling behaving as $\frac{1}{g_4^2}\sim \frac{x_{6,n+1}-x_{6,n}}{g_s\sqrt{\alpha^\prime}}$. The number of supersymmetries preserved is maintained if additional orthogonal (flavour) D6 branes extended in the $(\mathbb{R}^{1,3}, x_7,x_8,x_9)$ directions are added to the system, as we will do in section \ref{IIAwithD6}. The theory is  conformal if the number of flavours at each $[x_{6,n},x_{6,n+1}]$ interval is equal to twice the number of colours at the same interval. This theory is described holographically in Type IIA string theory by the class of Gaiotto-Maldacena geometries \cite{Gaiotto:2009gz}.

In the particular D4-NS5 system included in the intersection in Table \ref{Table:branesinmasslessIIA1} the D4-branes are stretched periodically between NS5-branes, that are positioned along the circular $\psi$-direction. This subsystem is depicted in Table \ref{Table:D4NS5system}.
\begin{table}[http!]
\renewcommand{\arraystretch}{1}
\begin{center}
\scalebox{1}[1]{
\begin{tabular}{c  c cc  c|| c  c  c |c c c}
 branes & $t$ & $x^1$ & $x^2$ & $x^3$ & $y$ & $z$ & $\psi$ & $r$ & $\theta^1$ & $\theta^2$ \\
\hline \hline
$\mrm{D}4$ & $\times$ & $\times$ & $\times$ & $\times$ & $-$ & $-$ & $\times$ & $-$ & $-$ & $-$ \\
$\mrm{NS}5$ & $\times$ & $\times$ & $\times$ & $\times$ & $\times$ & $\times$ & $-$ & $-$ & $-$ & $-$ \\
\end{tabular}
}
\caption{BPS/4 intersection describing the D4-NS5 bound state considered in section \ref{branesystemIIA}. The D4-branes extend along the $\psi$ circular direction, along which the NS5-branes are located. The system supports a 4d $\ma N=2$ CFT  \cite{Witten:1997sc}.} \label{Table:D4NS5system}
\end{center}
\end{table}
 As shown in \cite{Witten:1997sc,Fayyazuddin:1999zu,Alishahiha:1999ds},
the $\ma N=2$ CFT living in this brane system is a $\mathbb{Z}_n$ orbifold of $\ma N=4$ SYM, and it is holographically dual to a specific Abelian T-dual of AdS$_5\times S^5/\mathbb{Z}_n$.  Following \cite{Oz:1999qd}, we show next that this solution arises as the near horizon geometry of the intersecting D4-NS5 system depicted in Table \ref{Table:D4NS5system}, where the NS5-branes are taken to be smeared along the $\psi$-direction.

We start by writing down the brane background,
\begin{equation}
\label{brane_metric_D4NS5}
\begin{split}
d s_{10}^2 &= H_{\mathrm{D}4}^{-1/2}  ds^2_{\mathbb{R}^{1,3}} + H_{\mathrm{D}4}^{1/2} \left(dy^2+dz^2\right)+H_{\mathrm{NS}5}H_{\mathrm{D}4}^{-1/2}\,d\psi^2+H_{\mathrm{NS}5}H_{\mathrm{D}4}^{1/2}\bigl(dr^2+r^2ds^2_{\tilde S^2}\bigr)\,,\\
H_{(3)} &=\partial_r H_{\mathrm{NS}5}\,r^2\,d\psi \wedge\text{vol}_{ \tilde S^2}\,,\qquad e^{\Phi}= H_{\mathrm{NS}5}^{1/2}H_{\mathrm{D}4}^{-1/4}\,,\\
F_{(4)}&=\partial_r H_{\mathrm{D}4}\,r^2\,  dy\wedge dz \wedge \text{vol}_{ \tilde S^2}
+H_{\mathrm{NS}5}\partial_y H_{\mathrm{D}4}\,r^2\,  dz\wedge d r \wedge \text{vol}_{ \tilde S^2}
-H_{\mathrm{NS}5}\partial_z H_{\mathrm{D}4}\,r^2\,  dy\wedge d r \wedge \text{vol}_{ \tilde S^2}\,,
\end{split}
\end{equation}
where $\mathbb{R}^{1,3}$ is the common worldvolume to the D4-NS5 branes. The D4-branes are completely localised in their transverse space, such that $H_{\mathrm{D}4}=H_{\mathrm{D}4}(y,z,r)$, while the NS5-branes are smeared in $\psi$, and $H_{\mathrm{NS}5}=H_{\mathrm{NS}5}(r)$. As already mentioned,
the equations of motion and Bianchi identities for this brane background are the same already given by  \eqref{EOM_NS5D4}, namely
\begin{equation}
 \nabla^2_{\mathbb{R}^3_r} H_{\mathrm{D}4}+H_{\mathrm{NS}5}\nabla^2_{\mathbb{R}^2_{(y,z)}} H_{\mathrm{D}4}=0\qquad \text{and} \qquad  \nabla^2_{\mathbb{R}^3_r} H_{\mathrm{NS}5}=0\,.
\end{equation}
We can now consider the semi-localised solution with harmonic functions \cite{Youm:1999ti,Fayyazuddin:1999zu,Loewy:1999mn},
\begin{equation}\label{semilocalizedNS5D4}
 H_{\mathrm{D}4}=1+\frac{q_{\mathrm{D}4}}{(y^2+z^2+4 q_{\mathrm{NS}5}r)^2}\qquad \text{and}\qquad H_{\mathrm{NS}5}=\frac{q_{\mathrm{NS}5}}{r}\,,
\end{equation}
and introduce the following new coordinates $(\mu,\alpha,\phi)$ \cite{Oz:1999qd},
\begin{equation}\label{AdS5coord}
   y= \mu \sin \alpha\cos\phi\,,\qquad z= \mu \sin \alpha\sin\phi\qquad \text{and}\qquad r= 4^{-1}\,q_{\mathrm{NS}5}^{-1}\, \mu^2\cos^2\alpha\,,
\end{equation}
with $\mu>0$, $\alpha \in [0,\frac{\pi}{2}]$ and $\phi$ the angular polar coordinate within the $\mathbb{R}^2_{(y,z)}$ plane. In these new coordinates one can easily extract a near horizon limit by taking $\mu\rightarrow 0$, obtaining \cite{Oz:1999qd}
\begin{equation}
\label{brane_metric_D4NS5_nearhorizon}
\begin{split}
d s_{10}^2 &= q_{\mathrm{D}4}^{1/2} ds^2_{\text{AdS}_5}+q_{\mathrm{D}4}^{1/2}\left[d\alpha^2+s^2 d\phi^2+4q_{\mathrm{NS}5}^2q_{\mathrm{D}4}^{-1}c^{-2}d\psi^2+4^{-1}c^2ds^2_{\tilde S^2}  \right]\,,\\
H_{(3)} &=-q_{\mathrm{NS}5}\,d\psi \wedge\text{vol}_{ \tilde S^2}\,,\qquad e^{\Phi}= 2q_{\mathrm{NS}5}q_{\mathrm{D}4}^{-1/4}\,c^{-1}\,,\\
F_{(4)}&=2^{-1}q_{\mathrm{NS}5}^{-1}q_{\mathrm{D}4}\,c^3s\, d\phi\wedge d \alpha \wedge \text{vol}_{ \tilde S^2}\,,
\end{split}
\end{equation}
where $ds^2_{\text{AdS}_5}=q_{\mathrm{D}4}^{-1}\,\mu^2ds^2_{\mathbb{R}^{1,3}}+\frac{d\mu^2}{\mu^2}$ and, for simplicity of notation, $s=\sin\alpha,\,c=\cos\alpha$. This background is a Gaiotto-Maldacena geometry, with the $\mrm{SU}(2)\times \mrm U(1)$ R-symmetry  realised as rotations of the  $\tilde S^2$ and the $S^1_\phi$, respectively. In turn, $S^1_\psi$ is an extra circle that lives in the 2d Riemann surface contained in the transverse space. This has been discussed in 
 \cite{Lozano:2016kum}. This background is T-dual to the AdS$_5\times S^5/\mathbb{Z}_n$ solution of Type IIB supergravity, and it is therefore holographically dual to a $\mathbb{Z}_n$ orbifold of 4d $\ma N=4$ SYM. Indeed, applying Buscher's rules in the parametrisation
\begin{equation}
ds^2_{S^5/\mathbb{Z}_n}=d\alpha^2+\sin^2{\alpha}d\phi^2+\cos^2{\alpha}ds^2_{S^3/\mathbb{Z}_n},
\end{equation}
and
\begin{equation}
ds^2_{S^3/\mathbb{Z}_n}=\frac14 \Bigl[ \Bigl(\frac{d\psi}{n}+\cos{\xi}d\eta\Bigr)^2+d\xi^2+\sin^2{\xi}d\eta^2\Bigr],
\end{equation}
with $\psi$ the T-duality direction, the solution (\ref{brane_metric_D4NS5_nearhorizon}) is reproduced, with $n=q_{\text{NS}5}/\sqrt{q_{\text{D}4}}$.
It will be useful to recall in the following sections that this near horizon limit is not manifest in the coordinates of the brane background, but requires the non-linear change of coordinates given by \eqref{AdS5coord}.
 
 \subsection{F1-D2-D4$'$-NS5$'$ line defects within AdS$_5$}\label{AdS2defectAdS5}

Let us now come back to the more general situation in which we have F1-D2-D4$'$-NS5$'$ defect branes ending on the D4-NS5 system, and apply the same logic above. In this case we are interested in the $\rho \rightarrow 0$ limit of \eqref{stringsol}, and therefore in the $\ma N=4$ $\mrm{AdS}_2\times S^2\times \tilde S^2 \times \mathbb{R}^2\times S^1 $ backgrounds fibered over an interval defined by \eqref{brane_metric_D4NS5D2F1D4_nh}. 
These solutions have the crucial property that the backreaction of the F1-D2-D4$'$-NS5$'$ branes on the D4-NS5 system modifies only the 4d worldvolume space of the D4-NS5 solution, keeping intact its $\mathbb{R}^2_{(y,z)}\times \mathbb{R}^3_r$ transverse space. This follows from the fact that the equations of motion associated to the F1-D2-D4$'$-NS5$'$ branes, given by \eqref{D4NS5D2F1EOM}, and those of the D4-NS5 system, given by \eqref{EOM_NS5D4}, are completely independent. 

This implies, among other things, that the semi-localised solution specified by \eqref{semilocalizedNS5D4} for the D4-NS5 bound state is still a solution. The
presence of the defect branes breaks however the 
 isometries of the 4d worldvolume of the D4-NS5 intersection, turning it into an  $\mrm{AdS}_2\times S^2$ backreacted geometry. In this case the change of coordinates defined in \eqref{AdS5coord} allows one to extract an asymptotically locally AdS$_5$ geometry in the $\mu \rightarrow 0$ limit, given by
\begin{equation}
\label{AdS2defect}
\begin{split}
d s_{10}^2 = q_{\mathrm{D}2}^{1/2}&q_{\mathrm{F}1}^{-1/2}q_{\mathrm{D}4}^{1/2}\overbrace{\left[q_{\mathrm{D}4}^{-1}\,q_{\mathrm{D}2}\,q_{\mathrm{F}1}\mu^2\left(ds^2_{\text{AdS}_2}+ds^2_{S^2}\right)+\frac{d\mu^2}{\mu^2}   \right]}^{\text{locally}\,\,\, \text{AdS}_5\,\,\, \text{geometry}}\\ +&q_{\mathrm{D}2}^{1/2}q_{\mathrm{F}1}^{-1/2}q_{\mathrm{D}4}^{1/2}\left[d\alpha^2+s^2 d\phi^2+4q_{\mathrm{D}2}^{-1}q_{\mathrm{F}1}q_{\mathrm{NS}5}^2q_{\mathrm{D}4}^{-1}c^{-2}d\psi^2+4^{-1}c^2ds^2_{S^2}  \right]\,,
\end{split}
\end{equation}
with $s=\sin\alpha$ and $c=\cos\alpha$. One can see in \eqref{AdS2defect} that the internal manifold is the same one of the 
$\mrm{AdS}_5$ vacuum written in \eqref{brane_metric_D4NS5_nearhorizon}. 
In this new system of coordinates the fluxes and the dilaton take the form,
\begin{equation}
\label{fluxes_D4NS5D2F1D4_nh_defectcoord}
\begin{aligned}
 e^{\Phi}=&2q_{\mathrm{D}2}^{3/4} q_{\mathrm{F}1}^{-3/4}q_{\mathrm{NS}5}q_{\mathrm{D}4}^{-1/4}\,c^{-1}\,,\\
  H_{(3)} &= -q_{\mathrm{D}2}\left(\tilde s\,\text{vol}_{\text{AdS}_2}+\tilde c\,\text{vol}_{S^2}\right)\wedge \left( s\,d\mu+\mu c\,d\alpha \right)-q_{\mathrm{D}2}\mu s\left(\tilde c\,\text{vol}_{\text{AdS}_2}-\tilde s\,\text{vol}_{S^2}\right)\wedge d\phi\\
  &-q_{\text{NS}5}d\psi \wedge\text{vol}_{ \tilde S^2}\,,\\
F_{(4)}&=q_{\mathrm{F}1}\left(\tilde c\,\text{vol}_{\text{AdS}_2}-\tilde s\,\text{vol}_{S^2}\right)\wedge \left( s\,d\mu+\mu c\,d\alpha \right)\wedge d\psi\\
&-q_{\mathrm{F}1}\mu s\left(\tilde s\,\text{vol}_{\text{AdS}_2}+\tilde c\,\text{vol}_{S^2}\right)\wedge d\phi\wedge d\psi+2^{-1}q_{\mathrm{NS}5}^{-1}q_{\mathrm{D}4}\,c^3s\, d\phi\wedge d \alpha \wedge \text{vol}_{ \tilde S^2}\,,\\
\end{aligned}
\end{equation}
with $\tilde s=\sin\phi$ and $\tilde c=\cos\phi$.

This analysis shows that starting with the general brane intersection specified by the solutions  \eqref{brane_metric_D4NS5D2F1D4} and taking the particular profiles \eqref{stringsol} for the F1-D2-D4$'$-NS5$'$ branes and the semi-localised profile \eqref{semilocalizedNS5D4} for the D4-NS5 system, two interesting regimes emerge. The first regime is when $\rho \rightarrow 0$. In this case the defect branes are resolved into a fully backreacted $\mrm{AdS}_2\times S^2$ geometry within the 4d worldvolume of the D4-NS5 bound state, making manifest the breaking of its isometries. The second regime becomes manifest in the system of coordinates introduced in  \eqref{AdS5coord}, where besides the $\rho \rightarrow 0$ limit one takes the limit $\mu \rightarrow 0$, which allows one to approach the origin of the $\mathbb{R}^2_{(y,z)}$ plane.
In this regime the metric is split into a 5d ``external'' part, reproducing locally an $\mrm{AdS}_5$ geometry, and a 5d internal part, which, as we have mentioned, is shared with the internal part of the AdS$_5$ vacuum geometry found in \eqref{brane_metric_D4NS5_nearhorizon}. The isometries of the $\mrm{AdS}_5$ vacuum are however broken by the background fluxes, as shown by their expressions in \eqref{fluxes_D4NS5D2F1D4_nh_defectcoord}. The extra terms show that a 5d observer placed at $\mu \rightarrow 0$  feels the global charges of the defect branes, which backreact into a geometry described by a 5d curved domain wall with $\mrm{AdS}_2 \times S^2$ slicings, that is only locally AdS$_5$. Note that the presence of the extra terms in the fluxes forbids as well for any supersymmetric enhancement to  the (four dimensional) $\ma N=2$ supersymmetry of the AdS$_5$ solution \eqref{brane_metric_D4NS5_nearhorizon}.
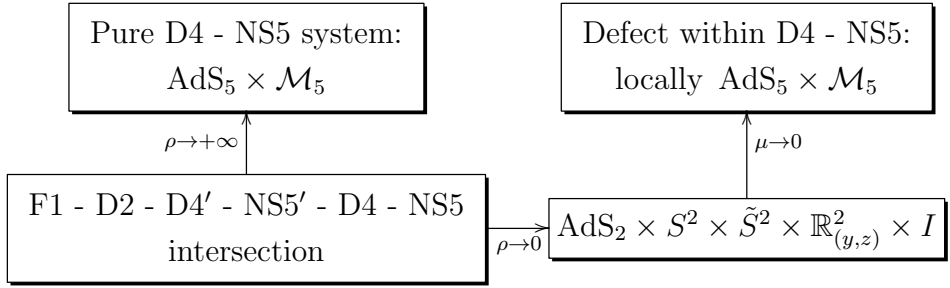
\begin{figure}[http!]
\begin{center}
\scalebox{1}[1]{\xymatrix{*+[F-,]{\begin{array}{c}\textrm{Pure D4 - NS5 system:}\\ \textrm{AdS}_{5}\times \ma M_5\end{array}} & *+[F-,]{\begin{array}{c}  \textrm{Defect within D4 - NS5:}\\ \textrm{locally\,\,\,} \textrm{AdS}_5\times \ma M_5 \end{array}} \\ 
*+[F-,]{\begin{array}{c} \textrm{F1 - D2 - D4$'$ - NS5$'$ - D4 - NS5}\\ \textrm{intersection}\end{array}} \ar[r]_{\hspace{8mm}\rho \rightarrow 0\,\,\,\,}\ar[u]^{\rho \rightarrow +\infty}
 &  *+[F-,]{\textrm{AdS}_{2}\times S^2 \times \tilde S^2 \times \mathbb{R}^2_{(y,z)} \times I}\ar[u]_{\mu \rightarrow 0}
 }}
\end{center}
\caption{The $\rho\rightarrow +\infty$ and $\rho\rightarrow 0$ limits of the F1-D2-D4$'$-NS5$'$-D4-NS5 brane intersection of Table \ref{Table:branesinmasslessIIA1}, with its defect structure. The $\rho\rightarrow +\infty$ limit zooms out the F1-D2-D4$'$-NS5$'$ branes, leaving behind an  $\textrm{AdS}_{5}\times \ma M_5$ solution associated to the D4-NS5 branes. The $\rho\rightarrow 0$ limit produces an AdS$_2$ near horizon geometry which, in turn, approaches asymptotically the $\textrm{AdS}_{5}\times \ma M_5$ geometry in the $\mu\rightarrow 0$ limit, allowing one to interpret the F1-D2-D4$'$-NS5$'$ branes as describing a defect within the 4d SCFT associated to the D4-NS5 system.}
\label{fig:defectInterpretation} 
\end{figure}

Our construction thus realises in Type IIA supergravity a conformal line defect in the 4d $\ma N=2$ SCFT that results by orbifolding the 4d $\ma N=4$ SYM CFT by $\mathbb{Z}_n$ (that is, the field theory dual to the AdS$_5$ solution in \eqref{brane_metric_D4NS5_nearhorizon}). The defect is described by a superconformal quantum mechanics that is holographically dual to an $\mrm{AdS}_2$ geometry with $\ma N=4$ supersymmetries (in one dimension).  In the limit in which this solution asymptotes to AdS$_5$ the defect, as seen from a 5d observer, occurs as an angular wedge located at the conformal boundary of $\mrm{AdS}_2$. One can see this explicitly rewriting the locally $\mrm{AdS}_5$ part of the background \eqref{AdS2defect} as,
\begin{equation}
 ds^2_5 \sim f^{-2} \left(-dt^2+d\tilde \rho^2+\tilde \rho^2 ds^2_{S^2}+ \tilde \rho^2 d\lambda^2  \right)\,,
\end{equation}
where $f^{-2}=\mu^2 \tilde \rho^{-2}$, $d\lambda=\mu^{-2}d\mu$ and $\tilde \rho$ parametrises the radial direction in AdS$_2$ in Poincar\'e coordinates. From the above expression one can see that the metric in the $(\tilde \rho, \lambda)$ plane develops a conical defect at $\tilde \rho=0$. This fixes the locus of the defect and allows one to interpret the $\mu$ coordinate as an angular coordinate parametrising the wedge in which a 5d observer probes the defect geometry.

Figure  \ref{fig:defectInterpretation} contains a summary of the two limits of the solution \eqref{brane_metric_D4NS5D2F1D4} describing the F1-D2-D4$'$-NS5$'$-D4-NS5 brane intersection of Table \ref{Table:branesinmasslessIIA1}, studied in this and the previous subsections.

\section{Line defects within D3 branes}\label{IIBpart}

In this section we turn to the Type IIB realisation of our previous constructions, where the main features already discussed become more transparent.

In so doing we construct a new class of AdS$_2$ solutions to Type IIB supergravity with $\ma N=4$ supersymmetry, and show that a solution in this class finds an interesting line defect interpretation within AdS$_5\times S^5/\mathbb{Z}_n$. The key observation, already noted in subsection \ref{AdS5semilocalD4NS5},  is that the AdS$_5$ solution that arises  far away from the Type IIA defects is the T-dual of AdS$_5\times S^5/\mathbb{Z}_n$. This solution was associated to a semi-localised intersection of D4 and NS5 branes. The T-duality takes place along the $\psi$ circular direction on which the D4 branes are stretched, giving D3 branes. In turn, the NS5 branes become KK-monopoles, giving rise to a 
 foliation of the circle and the emergence of the Lens space $S^3/\mathbb{Z}_n$. When the KK charge is one, we recover the round $S^5$ and the isotropic D3 brane.

We start recalling the main features of the semi-localised D3-KK system that gives rise to AdS$_5\times S^5/\mathbb{Z}_n$ close to the horizon. We then analyse in detail the T-dual realisation of the F1-D2-D4$'$-NS5$'$-D4-NS5 intersection discussed in subsection \ref{branesystemIIA}. This becomes a bound state of F1-D1-D5-NS5 branes ending on a D3-KK intersection. We provide the full brane solution and the near horizon limit of this system. The latter gives rise to a new class of $\ma N=4$ $\mrm{AdS}_2\times S^2\times  S^2 \times \mathbb{R}^2\times S^1$ geometries foliated over a line, solutions of Type IIB supergravity. We show that a suitable prescription for the distribution of charges of the D3-KK system produces a solution within this class that asymptotes locally to the AdS$_5\times S^5/\mathbb{Z}_n$ vacuum, thus allowing us to interpret this solution as describing a line defect CFT within $\ma N=4$ SYM modded by $\mathbb{Z}_n$. In the particular case in which no KK-monopoles are present the solution describes a line defect CFT within 4d $\ma N=4$ SYM, that preserves 1/4 of the supersymmetries.

\subsection{AdS$_5$ vacua from the D3-KK system}\label{D3KKsystem}

In this subsection we recall the brane solution that describes D3 branes intersected with KK monopoles, as well as the emergence of AdS$_5$ close to the horizon. We follow \cite{Cvetic:2000cj}, where the change of coordinates reproducing the AdS$_5$ vacuum was obtained. T-dualising the D4-NS5 brane system depicted in Table \ref{Table:D4NS5system} one obtains the D3-KK intersection shown in Table \ref{Table:D3KKsystem}.
\begin{table}[http!]
\renewcommand{\arraystretch}{1}
\begin{center}
\scalebox{1}[1]{
\begin{tabular}{c c cc  c|| c  c  |c c c c}
 branes & $t$ & $x^1$ & $x^2$ & $x^3$ & $y$ & $z$ & $\psi$ & $r$ & $\theta^1$ & $\theta^2$ \\
\hline \hline
$\mrm{D}3$ & $\times$ & $\times$ & $\times$ & $\times$ & $-$ & $-$ & $-$ & $-$ & $-$ & $-$ \\
$\mrm{KK}$ & $\times$ & $\times$ & $\times$ & $\times$ & $\times$ & $\times$ & $\mrm{ISO}$ & $-$ & $-$ & $-$ \\
\end{tabular}
}
\caption{Brane system corresponding to the D3-KK intersection. The $S^5$ in the near horizon of the D3 branes is broken into the Lens space $S^5/\mathbb{Z}_n$. T-duality along the $\psi$ direction reproduces back the D4-NS5 system depicted in Table \ref{Table:D4NS5system}.} \label{Table:D3KKsystem}
\end{center}
\end{table}
The background associated to this system is given by
\begin{equation}
\label{brane_metric_D3KK}
\begin{split}
d s_{10}^2 &= H_{\mathrm{D}3}^{-1/2}  ds^2_{\mathbb{R}^{1,3}} + H_{\mathrm{D}3}^{1/2} \left(dy^2+dz^2\right)+H_{\mathrm{D}3}^{1/2}\,\left(H_{\text{KK}}^{-1}\left(d\psi+q_{\text{KK}}\omega \right)^2+H_{\text{KK}}\left(dr^2+r^2ds^2_{\tilde S^2}  \right) \right)\,,\\
F_{(5)}&= \partial_r H_{\mathrm{D}3}^{-1}\,\text{vol}_{ \mathbb{R}^{1,3}}\wedge dr+\partial_y H_{\mathrm{D}3}^{-1}\,\text{vol}_{ \mathbb{R}^{1,3}}\wedge dy+\partial_z H_{\mathrm{D}3}^{-1}\,\text{vol}_{ \mathbb{R}^{1,3}}\wedge dz\\
&-r^2\partial_r H_{\mathrm{D}3} dy \wedge dz \wedge d\psi \wedge \text{vol}_{ \tilde S^2}- H_{\mathrm{KK}}r^2\partial_z H_{\mathrm{D}3} dy \wedge d\psi \wedge dr \wedge\text{vol}_{ \tilde S^2}\\
&+ H_{\mathrm{KK}}r^2\partial_y H_{\mathrm{D}3} dz \wedge d\psi \wedge dr \wedge\text{vol}_{ \tilde S^2}\,,
\end{split}
\end{equation}
where $d\omega=\text{vol}_{\tilde S^2}$ and $q_{\text{KK}}$ is the KK monopole charge. The function $H_{\text{KK}}$ is defined over the 3d space $\mathbb{R}^3_r$, while $H_{\mathrm{D}3}$ describes the geometry of the D3 branes and is fully localised in the transverse manifold $\mathbb{R}^2_{(y,z)}\times\mathbb{R}^3_r$, such that $H_{\mathrm{D}3}=H_{\mathrm{D}3}(y,z,r)$. 
Plugging in the Ansatz \eqref{brane_metric_D3KK} into the equations of motion and Bianchi identities of Type IIB supergravity one obtains the equations
\begin{equation}
 \nabla^2_{\mathbb{R}^3_r} H_{\mathrm{D}3}+H_{\text{KK}}\nabla^2_{\mathbb{R}^2_{(y,z)}} H_{\mathrm{D}3}=0\qquad \text{with} \qquad  H_{\mathrm{KK}}=\frac{q_{\text{KK}}}{r}\,. \label{D3-KKeqs}
\end{equation}
As for the D4-NS5 system, one can consider the semi-localised solution \cite{Youm:1999ti},
\begin{equation}\label{semilocalizedD3KK}
 H_{\mathrm{D}3}=1+\frac{q_{\mathrm{D}3}}{(y^2+z^2+4 q_{\text{KK}}r)^2}\qquad \text{and}\qquad H_{\text{KK}}=\frac{q_{\text{KK}}}{r}\,,
\end{equation}
and one can introduce the new coordinates $(\mu,\alpha,\phi)$ \cite{Oz:1999qd,Cvetic:2000cj},
\begin{equation}\label{AdS5coordIIB}
   y= \mu \sin \alpha\cos\phi\,,\qquad z= \mu \sin \alpha\sin\phi\qquad \text{and}\qquad r= 4^{-1}\,q_{\text{KK}}^{-1}\, \mu^2\cos^2\alpha\,.
\end{equation}
In these coordinates the D3-KK background takes the form \cite{Cvetic:2000cj}
\begin{equation}
\begin{split}
 d s_{10}^2 &= H_{\mathrm{D}3}^{-1/2}  ds^2_{\mathbb{R}^{1,3}} + H_{\mathrm{D}3}^{1/2}\,\left(d\mu^2+\mu^2 ds^2_{ M_5}  \right) \,,\\
 ds^2_{M5}&=d\alpha^2+s^2d\phi^2+c^2ds^2_{S^3/\mathbb{Z}_n}\,,\\
 H_{\mathrm{D}3}&=1+\frac{q_{\mathrm{D}3}}{\mu^4}\,,
 \end{split}
\end{equation}
where $s=\sin\alpha$, $c=\cos \alpha$ and the orbifolded 3-sphere is written as
\begin{equation}
d s^2_{S^3/\mathbb{Z}_n}=\frac14\left[\left(\frac{d\psi}{n} +\omega \right)^2+ds^2_{\tilde S^2}  \right]\,.
\end{equation}
The internal manifold $M_5$ is thus a foliation of $S^1_\phi\times S^3/\mathbb{Z}_n$ over an interval, parametrised by the coordinate $\alpha$, with $n=q_{\text{KK}}$. This builds a $S^5/\mathbb{Z}_n$ space. Indeed, in 
these coordinates the near horizon limit is realised by taking $\mu\rightarrow 0$, giving rise to the AdS$_5\times S^5/\mathbb{Z}_n$ geometry \cite{Cvetic:2000cj},
\begin{equation}
\label{brane_metric_D3KK_nearhorizon}
\begin{split}
d s_{10}^2 &= q_{\mathrm{D}3}^{1/2} ds^2_{\text{AdS}_5}+q_{\mathrm{D}3}^{1/2}ds^2_{S^5/\mathbb{Z}_n}\,,\\
F_{(5)}&= 4q_{\mathrm{D}3} \left(1+\star_{(10)}\right)\, \text{vol}_{\text{AdS}_5}\,,
\end{split}
\end{equation}
with $ds^2_{\text{AdS}_5}=q_{\mathrm{D}3}^{-1}\,\mu^2ds^2_{\mathbb{R}^{1,3}}+\frac{d\mu^2}{\mu^2}$. 
An obvious interesting case is when $n=1$, where the D3 branes become isotropic and the metric over the $M_5$ describes a round $S^5$. In this situation there is a supersymmetry enhancement of the brane set-up to 16 real supercharges.

\subsection{The D1-F1-D5-NS5-D3-KK  brane set-up}

We consider now the Type IIA brane set-up depicted in  Table \ref{Table:branesinmasslessIIA1}, that we T-dualise along the $\psi$ circular direction. The F1-D2-D4$'$-NS5$'$ defect branes become a D1-F1-D5-NS5 brane system localised within the common worldvolume of the D3-KK branes.
\begin{table}[http!]
\renewcommand{\arraystretch}{1}
\begin{center}
\scalebox{1}[1]{
\begin{tabular}{c c cc  c|| c  c  |c c c c}
 branes & $t$ & $\rho$ & $\varphi^1$ & $\varphi^2$ & $y$ & $z$ & $\psi$ & $r$ & $\theta^1$ & $\theta^2$ \\
\hline \hline
$\mrm{D}3$ & $\times$ & $\times$ & $\times$ & $\times$ & $-$ & $-$ & $-$ & $-$ & $-$ & $-$ \\
$\mrm{KK}$ & $\times$ & $\times$ & $\times$ & $\times$ & $\times$ & $\times$ & $\mrm{ISO}$ & $-$ & $-$ & $-$ \\
$\mrm{D}1$ & $\times$ & $-$ & $-$ & $-$ & $\times$ & $-$ & $-$ & $-$ & $-$ & $-$ \\
$\mrm{F}1$ & $\times$ & $-$ & $-$ & $-$ & $-$ & $\times$ & $-$ & $-$ & $-$ & $-$ \\
$\mrm{D}5$ & $\times$ & $-$ & $-$ & $-$ & $\times$ & $-$ & $\times$ & $\times$ & $\times$ & $\times$ \\
$\mrm{NS}5$ & $\times$ & $-$ & $-$ & $-$ & $-$ & $\times$ & $\times$ & $\times$ & $\times$ & $\times$ \\
\end{tabular}
}
\caption{BPS/8 intersection describing D1-F1-D5-NS5 branes ending on the D3-KK system.} \label{Table:D1-F1-D5-NS5-D3-KKsystem}
\end{center}
\end{table}
The metric associated to this brane intersection has the general form
\begin{equation}
\label{brane_metric_D1-F1-D5-NS5-D3-KK}
\begin{split}
d s_{10}^2 &= H_{\mathrm{D}3}^{-1/2}  \left[- H_{\mathrm{D}1}^{-1/2} H_{\mathrm{F}1}^{-1} H_{\mathrm{D}5}^{-1/2}dt^2 +H_{\mathrm{D}1}^{1/2}  H_{\mathrm{D}5}^{1/2}H_{\mathrm{NS}5}\left(d\rho^2+\rho^2ds^2_{S^2} \right)  \right] \\
&+ H_{\mathrm{D}3}^{1/2} \left[H_{\mathrm{D}1}^{-1/2}  H_{\mathrm{D}5}^{-1/2}H_{\mathrm{NS}5}dy^2+H_{\mathrm{D}1}^{1/2} H_{\mathrm{F}1}^{-1} H_{\mathrm{D}5}^{1/2}dz^2\right]\\
&+H_{\mathrm{D}3}^{1/2}H_{\mathrm{D}1}^{1/2}  H_{\mathrm{D}5}^{-1/2}\,\left[H_{\text{KK}}^{-1}\left(d\psi+q_{\text{KK}}\omega \right)^2+H_{\text{KK}}\left(dr^2+r^2ds^2_{\tilde S^2}  \right) \right]\,,\\
e^{\Phi}&= H_{\mathrm{NS}5}^{1/2}H_{\mathrm{D}5}^{-1/2}H_{\mathrm{F}1}^{-1/2}H_{\mathrm{D}1}^{1/2}\,,
\end{split}
\end{equation}
where $d\omega=\text{vol}_{\tilde S^2}$ and $q_{\text{KK}}$ is the KK monopole charge. As in the previous subsection, the 
function $H_{\text{KK}}$ is defined over the 3d space $\mathbb{R}^3_r$ and $H_{\mathrm{D}3}=H_{\mathrm{D}3}(y,z,r)$.
For the defect branes we take the charge distributions localised within the worldvolume of the D3 branes, namely $H_{\mathrm{D}1}(\rho)$, $H_{\mathrm{F}1}(\rho)$, $H_{\mathrm{NS}5}(\rho)$, $H_{\mathrm{D}5}(\rho)$. The fluxes corresponding to this charge distribution take the form,
\begin{equation}
\begin{split}\label{fluxes_D1-F1-D5-NS5-D3-KK}
H_{(3)} &= -\partial_\rho H_{\mathrm{F}1}^{-1}dt\wedge d\rho\wedge dz+ \partial_\rho H_{\mathrm{NS}5}\,\rho^2\,\text{vol}_{ S^2}\wedge dy\,,\\
F_{(3)}&=-\partial_\rho H_{\mathrm{D}1}^{-1}dt\wedge d\rho\wedge dy-\partial_\rho H_{\mathrm{D}5}\,\rho^2\,\text{vol}_{ S^2}\wedge dz\,,\\
F_{(5)}&=H_{\mathrm{D}5}H_{\mathrm{NS}5}\rho^2\partial_r H_{\mathrm{D}3}^{-1}\,dt\wedge d\rho\wedge \text{vol}_{ S^2}\wedge dr+H_{\mathrm{D}5}H_{\mathrm{NS}5}\rho^2\partial_y H_{\mathrm{D}3}^{-1}\,dt\wedge d\rho\wedge \text{vol}_{ S^2}\wedge dy\\
&+H_{\mathrm{D}5}H_{\mathrm{NS}5}\rho^2\partial_z H_{\mathrm{D}3}^{-1}\,dt\wedge d\rho\wedge \text{vol}_{ S^2}\wedge dz-r^2\partial_r H_{\mathrm{D}3} dy \wedge dz \wedge d\psi \wedge \text{vol}_{ \tilde S^2}\\
&-H_{\mathrm{F}1}H_{\mathrm{D}5}^{-1} H_{\mathrm{KK}}r^2\partial_z H_{\mathrm{D}3} dy \wedge d\psi \wedge dr \wedge\text{vol}_{ \tilde S^2}+H_{\mathrm{D}1}H_{\mathrm{NS}5}^{-1} H_{\mathrm{KK}}r^2\partial_y H_{\mathrm{D}3} dz \wedge d\psi \wedge dr \wedge\text{vol}_{ \tilde S^2}\,.
\end{split}
\end{equation}
As usual, the equations of motion and Bianchi identities decouple into two groups, one for the  D1-F1-D5-NS5 defect branes 
\begin{equation} \label{D1-F1-D5-NS5EOM}
\begin{split}
&\nabla^2_{\mathbb{R}^3_\rho} H_{\mathrm{D}1}=0 \qquad \text{with}\qquad  H_{\mathrm{NS}5}=H_{\mathrm{D}1}\,,\\
&\nabla^2_{\mathbb{R}^3_\rho} H_{\mathrm{F}1}=0 \qquad \text{with}\qquad  H_{\mathrm{D}5}=H_{\mathrm{F}1}\,,\\
\end{split}
\end{equation}
and one for the D3-KK system 
\begin{equation}\label{EOM_D3KK1}
 \nabla^2_{\mathbb{R}^3_r} H_{\mathrm{D}3}+H_{\mathrm{KK}}\nabla^2_{\mathbb{R}^2_{(y,z)}} H_{\mathrm{D}3}=0\qquad \text{with} \qquad  H_{\mathrm{KK}}=\frac{q_{\text{KK}}}{r}\,.
\end{equation}
We consider now the particular solution for the defect branes,
\begin{equation}\label{stringsolIIB}
  H_{\mathrm{D}1}=1+\frac{q_{\mathrm{D}1}}{\rho} \qquad \text{and} \qquad H_{\mathrm{F}1}=1+\frac{q_{\mathrm{F}1}}{\rho}\,,
\end{equation}
where $q_{\mathrm{D}1}$ and $q_{\mathrm{F}1}$ are integration constants related to the quantised charges of the respective branes.
The limit $\rho \rightarrow + \infty$ reproduces the situation in which the defect branes are taken far away from the D3-KK  branes. This is the brane solution studied in the previous subsection. In turn, the $\rho \rightarrow 0$ limit gives rise to a new class of  $\ma N=4$ $\mrm{AdS}_2$ backgrounds\footnote{In order to reproduce unitary AdS$_2$ at the horizon we rescaled the time as $t\rightarrow q_{\mathrm{D}1} q_{\mathrm{F}1}t$.} of Type IIB string theory,
\begin{equation}
\label{brane_metric_D1-F1-D5-NS5-D3-KK_nh}
\begin{aligned}
ds_{10}^2 &= q_{\mathrm{D}1}^{3/2}q_{\mathrm{F}1}^{1/2}    H_{\mathrm{D}3}^{-1/2}\left(ds^2_{\text{AdS}_2}+ds^2_{S^2}\right) +q_{\mathrm{D}1}^{1/2}q_{\mathrm{F}1}^{-1/2} H_{\mathrm{D}3}^{1/2}\,  \left( dy^2+dz^2 \right)\\
&+q_{\mathrm{D}1}^{1/2}q_{\mathrm{F}1}^{-1/2}H_{\mathrm{D}3}^{1/2}\,\left(H_{\text{KK}}^{-1}\left(d\psi+q_{\text{KK}}\omega \right)^2+H_{\text{KK}}\left(dr^2+r^2ds^2_{\tilde S^2}  \right) \right)\,,\\
 e^{\Phi}&=q_{\mathrm{D}1}q_{\mathrm{F}1}^{-1}   \,,\qquad H_{(3)} = -q_{\mathrm{D}1}\text{vol}_{\text{AdS}_2}\wedge dz-q_{\mathrm{D}1} \text{vol}_{  S^2}\wedge dy\,,\\
F_{(3)}&=-q_{\mathrm{F}1}\text{vol}_{\text{AdS}_2}\wedge dy+q_{\mathrm{F}1} \text{vol}_{  S^2}\wedge dz\,,\\
F_{(5)}&=q_{\mathrm{D}1}^{2}q_{\mathrm{F}1}^{2} \partial_r H_{\mathrm{D}3}^{-1}\,\text{vol}_{\text{AdS}_2}\wedge \text{vol}_{ S^2}\wedge dr+q_{\mathrm{D}1}^{2}q_{\mathrm{F}1}^{2} \partial_y H_{\mathrm{D}3}^{-1}\,\text{vol}_{\text{AdS}_2}\wedge \text{vol}_{ S^2}\wedge dy\\
&+q_{\mathrm{D}1}^{2}q_{\mathrm{F}1}^{2} \partial_z H_{\mathrm{D}3}^{-1}\,\text{vol}_{\text{AdS}_2}\wedge \text{vol}_{ S^2}\wedge dz-r^2\partial_r H_{\mathrm{D}3} dy \wedge dz \wedge d\psi \wedge \text{vol}_{ \tilde S^2}\\
&-H_{\mathrm{KK}}r^2\partial_z H_{\mathrm{D}3} dy \wedge d\psi \wedge dr \wedge\text{vol}_{ \tilde S^2}+ H_{\mathrm{KK}}r^2\partial_y H_{\mathrm{D}3} dz \wedge d\psi \wedge dr \wedge\text{vol}_{ \tilde S^2}\,,
\end{aligned}
\end{equation}
where $H_{\mathrm{D}3}$ solves the master equation \eqref{EOM_D3KK1}. These geometries represent a new class of solutions to Type IIB supergravity, described by  $\mrm{AdS}_2\times S^2\times \tilde S^2 \times \mathbb{R}^2\times S^1 $ foliations over a line. One can easily check that these backgrounds are related by T-duality along the $\psi$ direction to the AdS$_2$ solutions given by \eqref{brane_metric_D4NS5D2F1D4_nh}, describing the near horizon of the F1-D2-D4$'$-NS5$'$-D4-NS5 brane intersections in Type IIA discussed in section \ref{IIApart}.

\subsection{Defects within 4d SCFTs in Type IIB}

In this subsection we follow the same approach taken in subsection \ref{AdS2defectAdS5} to provide a defect interpretation for the backgrounds given by \eqref{brane_metric_D1-F1-D5-NS5-D3-KK_nh}. As in subsection \ref{AdS2defectAdS5}  the crucial property that allows to find such an interpretation is the decoupling between the dynamics of the D1-F1-D5-NS5 defect branes and that of the D3-KK system (in the sense that the equations of motion and Bianchi identities of the two groups of branes, given by \eqref{D1-F1-D5-NS5EOM} and \eqref{EOM_D3KK1}, are completely independent). Since we are searching for a possible completion within an AdS$_5$ vacuum, we choose the semi-localised solution for the D3-KK system considered in subsection \ref{D3KKsystem}, given by equation \ref{semilocalizedD3KK}.
Using then the $(\mu, \alpha, \phi)$ coordinates 
introduced in \eqref{AdS5coordIIB} the backgrounds \eqref{brane_metric_D1-F1-D5-NS5-D3-KK_nh}
 take  the form of a stack of D3 branes wrapping the $\mrm{AdS}_2\times S^2$ backreacted geometry. The D3 branes are  intersected with $n$ KK-monopoles that turn the $S^5$ transverse space into $S^5/\mathbb{Z}_n$,
\begin{equation}
\label{brane_metric_D1-F1-D5-NS5-D3-KK_defect}
\begin{aligned}
ds_{10}^2 &= q_{\mathrm{D}1}^{3/2}q_{\mathrm{F}1}^{1/2}    H_{\mathrm{D}3}^{-1/2}\left(ds^2_{\text{AdS}_2}+ds^2_{S^2}\right) +q_{\mathrm{D}1}^{1/2}q_{\mathrm{F}1}^{-1/2} H_{\mathrm{D}3}^{1/2}\,\left(d\mu^2+\mu^2 ds^2_{S^5/\mathbb{Z}_n}  \right) \,,\\
 ds^2_{S^5/\mathbb{Z}_n}&=d\alpha^2+s^2d\phi^2+c^2ds^2_{S^3/\mathbb{Z}_n}\,,\\
 H_{\mathrm{D}3}&=1+\frac{q_{\mathrm{D}3}}{\mu^4}\,.
\end{aligned}
\end{equation}
As we already observed in the Type IIA case, the space transverse to the D3 branes  is left untouched by the intersection with the  defect branes, whose presence is only manifest through the fully-backreacted $\mrm{AdS}_2\times S^2$ geometry, that curves the worldvolume of the D3 branes. Interestingly, in these coordinates the 
background admits a locally $\mrm{AdS}_5\times S^5/\mathbb{Z}_n$ geometry in the $\mu \rightarrow 0$ limit,
\begin{equation}
\begin{split}\label{defectIIB}
d s_{10}^2 &= q_{\mathrm{D}3}^{1/2}q_{\mathrm{D}1}^{1/2}q_{\mathrm{F}1}^{-1/2} \overbrace{\left[q_{\mathrm{D}3}^{-1}\,q_{\mathrm{D}1}\,q_{\mathrm{F}1}\mu^2\left(ds^2_{\text{AdS}_2}+ds^2_{S^2}\right)+\frac{d\mu^2}{\mu^2}   \right]}^{\text{locally}\,\,\, \text{AdS}_5\,\,\, \text{geometry}}\\
&+q_{\mathrm{D}3}^{1/2}q_{\mathrm{D}1}^{1/2}q_{\mathrm{F}1}^{-1/2} \left[d\alpha^2+s^2d\phi^2+c^2ds^2_{S^3/\mathbb{Z}_k} \right]\,,\\
e^{\Phi}&=q_{\mathrm{D}1}q_{\mathrm{F}1}^{-1}\,,\\
H_{(3)}& = -q_{\mathrm{D}1}\left(\tilde s \text{vol}_{\text{AdS}_2}+\tilde c \text{vol}_{  S^2} \right)\wedge \left(sd\mu+\mu c d\alpha \right)-q_{\mathrm{D}1}\mu s\left( \tilde c \text{vol}_{\text{AdS}_2}-\tilde s \text{vol}_{  S^2}\right) \wedge d\phi\,,\\
F_{(3)}& = q_{\mathrm{F}1}\left(-\tilde c \text{vol}_{\text{AdS}_2}+\tilde s \text{vol}_{  S^2} \right)\wedge \left(sd\mu+\mu c d\alpha \right)+q_{\mathrm{F}1}\mu s\left( \tilde s \text{vol}_{\text{AdS}_2}+\tilde c \text{vol}_{  S^2}\right) \wedge d\phi\,,\\
F_{(5)}&=4q_{\mathrm{F}1}^2q_{\mathrm{D}1}^2q_{\mathrm{D}3}^{-1}\mu^3\text{vol}_{\text{AdS}_2}\wedge \text{vol}_{S^2}\wedge d\mu-2^{-1}q_{\mathrm{D}3}q_{\text{KK}}^{-1}c^3sd\phi\wedge d\alpha\wedge d\psi \wedge \text{vol}_{S^2}\,,
\end{split}
\end{equation}
where $s=\sin\alpha$, $c=\cos \alpha$ and $\tilde s=\sin\phi$, $\tilde c=\cos \phi$.

In complete analogy with the analysis performed in section \ref{IIApart}, our analysis in this section shows that starting with the general brane intersection specified by the solutions  \eqref{brane_metric_D1-F1-D5-NS5-D3-KK} and taking the particular profiles \eqref{stringsolIIB} for the D1-F1-D5-NS5 branes and the semi-localised profile \eqref{semilocalizedD3KK} for the D3-KK system, two interesting regimes emerge. The first regime is when $\rho \rightarrow 0$. In this case the defect branes are resolved into a fully backreacted $\mrm{AdS}_2\times S^2$ geometry within the 4d worldvolume of the D3-KK bound state, making manifest the breaking of its isometries. The second regime becomes manifest in the system of coordinates introduced in  \eqref{AdS5coordIIB}, where besides the $\rho \rightarrow 0$ limit one takes the limit $\mu \rightarrow 0$, which allows one to approach the origin of the $\mathbb{R}^2_{(y,z)}$ plane.
In this regime the metric is split into a 5d ``external'' part, reproducing locally an $\mrm{AdS}_5$ geometry, and a 5d internal part, shared with the internal part of the AdS$_5\times S^5/\mathbb{Z}_n$ vacuum. The isometries of the $\mrm{AdS}_5$ vacuum are however broken by the background fluxes, as shown by their expressions in \eqref{defectIIB}. The extra terms show that a 5d observer placed at $\mu \rightarrow 0$  feels the global charges of the defect branes, which backreact into a geometry described by a 5d curved domain wall with $\mrm{AdS}_2 \times S^2$ slicings, that is only locally AdS$_5$. Note that the presence of the extra terms in the fluxes forbids as well any supersymmetric enhancement to  the (four dimensional) $\ma N=2$ supersymmetry of the AdS$_5$ solution. 

As in subsection 
 \ref{AdS2defectAdS5}, our construction realises a conformal line defect in 4d $\ma N=4$ SYM modded by $\mathbb{Z}_n$, this time in terms of D1-F1-D5-NS5 branes. The Type IIB realisation allows one however to study the interesting case in which these defect branes are introduced within 4d $\ma N=4$ SYM, breaking the supersymmetries to 1/4 BPS. In this case it should be possible to interpret the D5 and the F1 branes as realising the baryon vertex of 4d $\ma N=4$ SYM, and the AdS$_2$ solutions as describing the corresponding backreacted geometries. In the IR the gauge symmetry on the D3 branes would become global, turning them into flavour branes, with the D5 branes becoming the new colour branes. Note however that in the backreacted geometry there are as well D1 colour branes. These should find an interpretation in terms of instantons within the worldvolume of the D5 branes. The possibility of such an interpretation will become clearer after our field theory analysis in the next section.

\section{Back to Type IIA: AdS$_2$ solutions with D6 branes}\label{IIAwithD6}

In this section we generalise the F1-D2-D4$'$-NS5$'$-D4-NS5 brane intersection studied in section 2 to include D6 branes localised within the $\mathbb{R}^2_{(y,z)}$ plane. We will see that adding D6 branes challenges the construction of a solution with AdS$_5$ asymptotics. However, we will see that taking a simplified ansatz it is possible to construct an explicit quiver quantum mechanics that can be interpreted as describing D2-D4' baryon vertices within the 4d SCFT living in the D4-NS5-D6 brane intersection.

\subsection{Adding D6 branes to the brane set-up}\label{D6inclusion}

The brane set-up discussed in section \ref{branesystemIIA} can be extended by including D6 branes localised within the $\mathbb{R}^2_{(y,z)}$ plane. The extended brane set-up is depicted in Table \ref{Table:branesinmasslessIIA1D6}.
\begin{table}[http!]
\renewcommand{\arraystretch}{1}
\begin{center}
\scalebox{1}[1]{
\begin{tabular}{cc| cc c|| c  c  c| c c c}
 branes & $t$ & $\rho$ & $\varphi^1$ & $\varphi^{2}$ & $y$ & $z$ & $\psi$ & $r$ & $\theta^1$ & $\theta^2$ \\
\hline \hline
$\mrm{D}6$ & $\times$ & $\times$ & $\times$ & $\times$ & $-$ & $-$ & $-$ & $\times$ & $\times$ & $\times$ \\
$\mrm{D}4$ & $\times$ & $\times$ & $\times$ & $\times$ & $-$ & $-$ & $\times$ & $-$ & $-$ & $-$ \\
$\mrm{NS}5$ & $\times$ & $\times$ & $\times$ & $\times$ & $\times$ & $\times$ & $-$ & $-$ & $-$ & $-$ \\
$\mrm{D}2$ & $\times$ & $-$ & $-$ & $-$ & $\times$ & $-$ & $\times$ & $-$ & $-$ & $-$ \\
$\mrm{F}1$ & $\times$ & $-$ & $-$ & $-$ & $-$ & $\times$ & $-$ & $-$ & $-$ & $-$ \\
$\mrm{D}4'$ & $\times$ & $-$ & $-$ & $-$ & $\times$ & $-$ & $-$ & $\times$ & $\times$ & $\times$ \\
$\mrm{NS}5'$ & $\times$ & $-$ & $-$ & $-$ & $-$ & $\times$ & $\times$ & $\times$ & $\times$ & $\times$ \\
\end{tabular}
}
\caption{BPS/8 intersection describing F1-D2-D4$'$-NS5$'$ branes ending on a D4-NS5-D6 bound state. } \label{Table:branesinmasslessIIA1D6}
\end{center}
\end{table}
This generalisation does not imply any further breaking of supersymmetries. However, as we already mentioned, it complicates finding an eventual AdS$_5$ completion. 

For simplicity, we consider adding D6 branes localised in the $\mathbb{R}^2_{(y,z)}$ plane but smeared along the circular direction $\psi$. 
 The metric and dilaton take the form
\begin{equation}
\label{brane_metric_D6D4NS5D2F1D4}
\begin{split}
d s_{10}^2 &= H_{\mathrm{D}6}^{-1/2}H_{\mathrm{D}4}^{-1/2}  \left[-H_{\mathrm{F}1}^{-1}H_{\mathrm{D}2}^{-1/2} H_{\mathrm{D}4'}^{-1/2} \,dt^2+H_{\mathrm{D}2}^{1/2} H_{\mathrm{D}4'}^{1/2}H_{\mathrm{NS}5'} \bigl(d\rho^2+\rho^2ds^2_{S^2}\bigr) \right] \\
&+H_{\mathrm{D}6}^{1/2} H_{\mathrm{D}4}^{1/2} \left[H_{\mathrm{D}2}^{-1/2} H_{\mathrm{D}4'}^{-1/2}H_{\mathrm{NS}5'}\,dy^2+H_{\mathrm{F}1}^{-1} H_{\mathrm{D}2}^{1/2} H_{\mathrm{D}4'}^{1/2}\,dz^2 \right]\\
&+H_{\mathrm{D}6}^{1/2}H_{\mathrm{NS}5}H_{\mathrm{D}4}^{-1/2}H_{\mathrm{D}2}^{-1/2} H_{\mathrm{D}4'}^{1/2}\,d\psi^2+H_{\mathrm{D}6}^{-1/2}H_{\mathrm{NS}5}H_{\mathrm{D}4}^{1/2}H_{\mathrm{D}2}^{1/2} H_{\mathrm{D}4'}^{-1/2}\bigl(dr^2+r^2ds^2_{\tilde S^2}\bigr)\,,\\
e^{\Phi}&= H_{\mathrm{D}6}^{-3/4}H_{\mathrm{NS}5}^{1/2}H_{\mathrm{D}4}^{-1/4}H_{\mathrm{F}1}^{-1/2}H_{\mathrm{D}2}^{1/4}H_{\mathrm{D}4'}^{-1/4}H_{\mathrm{NS}5'}^{1/2},
\end{split}
\end{equation}
where $H_{\mathrm{D}6}=H_{\mathrm{D}6}(y,z)$.
The fluxes are
\begin{equation}
\begin{split}\label{fluxes_D6D4NS5D2F1D4}
H_{(3)} &= -\partial_\rho H_{\mathrm{F}1}^{-1}dt\wedge d\rho\wedge dz+ \partial_\rho H_{\mathrm{NS}5'}\,\rho^2\, \text{vol}_{  S^2}\wedge dy+ \partial_r H_{\mathrm{NS}5}\,r^2\,d\psi \wedge\text{vol}_{ \tilde S^2}\,,\\
F_{(2)}&=-H_{\mathrm{F}1}H_{\mathrm{D}2}^{-1}\partial_zH_{\mathrm{D}6}dy\wedge d\psi+H_{\mathrm{D}4'}H_{\mathrm{NS}5'}^{-1}\partial_yH_{\mathrm{D}6}dz\wedge d\psi\,,\\
F_{(4)}&=H_{\mathrm{D}6}\partial_\rho H_{\mathrm{D}2}^{-1}dt\wedge d\rho\wedge dy \wedge d\psi+ H_{\mathrm{D}6}\partial_\rho H_{\mathrm{D}4'}\,\rho^2\, \text{vol}_{  S^2}\wedge dz\wedge d\psi\\
&+H_{\mathrm{D}2}H_{\mathrm{NS}5'}^{-1}H_{\mathrm{NS}5}\partial_y H_{\mathrm{D}4}\,r^2\,  dz\wedge d r \wedge \text{vol}_{ \tilde S^2}
-H_{\mathrm{F}1}H_{\mathrm{D}4'}^{-1}H_{\mathrm{NS}5}\partial_z H_{\mathrm{D}4}\,r^2\,  dy\wedge d r \wedge \text{vol}_{ \tilde S^2}\\
&+H_{\mathrm{D}6}\partial_r H_{\mathrm{D}4}\,r^2\,  dy\wedge dz \wedge \text{vol}_{ \tilde S^2}\,,\\
F_{(6)}&=H_{\mathrm{D}4'}H_{\mathrm{NS}5'} \rho^2 \Bigl(\partial_y H_{\mathrm{D}4}^{-1} dy +\partial_z H_{\mathrm{D}4}^{-1} dz+\partial_r H_{\mathrm{D}4}^{-1} dr\Bigr)\wedge dt\wedge d\rho\wedge  \text{vol}_{  S^2}\wedge d\psi\\
&-H_{\mathrm{NS}5}H_{\mathrm{D}4}\partial_\rho H_{\mathrm{D}2} r^2\rho^2 \text{vol}_{  S^2}\wedge dz\wedge dr\wedge \text{vol}_{ \tilde S^2}-H_{\mathrm{D}4} H_{\mathrm{NS}5} r^2 \partial_\rho H_{\mathrm{D}4'}^{-1} dt\wedge d\rho\wedge dy\wedge dr \wedge  \text{vol}_{ \tilde S^2}\, . \\
\end{split}
\end{equation}
As in the case without D6 branes the equations of motion and Bianchi identities decouple into two groups, one associated to the F1-D2-D4$'$-NS5$'$ defect branes,
\begin{equation} \label{D4NS5D2F1EOM1}
\begin{split}
&\nabla^2_{\mathbb{R}^3_\rho} H_{\mathrm{D}2}=0 \qquad \text{with}\qquad  H_{\mathrm{D}4'}=H_{\mathrm{NS}5'}=H_{\mathrm{F}1}=H_{\mathrm{D}2}\,,\\
\end{split}
\end{equation}
and a second one associated to the D4-NS5-D6 brane system,
\begin{equation}\label{EOM_D6NS5D4}
 H_{\mathrm{D}6}\nabla^2_{\mathbb{R}^3_r} H_{\mathrm{D}4}+H_{\mathrm{NS}5}\nabla^2_{\mathbb{R}^2_{(y,z)}} H_{\mathrm{D}4}=0\,, \qquad  \nabla^2_{\mathbb{R}^3_r} H_{\mathrm{NS}5}=0\qquad \text{and}\qquad \nabla^2_{\mathbb{R}_{(y,z)}^2} H_{\mathrm{D}6}=0\,.
\end{equation}

In order to extract the AdS$_2$ near horizon geometry we make the choice 
\begin{equation}\label{stringsolD6}
  H_{\mathrm{D}2}=1+\frac{q_{\mathrm{D}2}}{\rho}\,,
\end{equation}
where $q_{\mathrm{D}2}$ is an integration constant related to the quantised charges of the defect branes. We then obtain upon taking the $\rho \rightarrow 0$ limit\footnote{In order to reproduce unitary AdS$_2$ at the horizon we rescaled the time as $t\rightarrow q_{\mathrm{D}2}^2t$.},
\begin{equation}
\label{brane_metric_D6D4NS5D2F1D4_nh}
\begin{aligned}
ds_{10}^2 &= q_{\mathrm{D}2}^{2}  H_{\mathrm{D}6}^{-1/2} H_{\mathrm{D}4}^{-1/2}\left(ds^2_{\text{AdS}_2}+ds^2_{S^2}\right) +H_{\mathrm{D}6}^{1/2}  H_{\mathrm{D}4}^{1/2}\, (dy^2+dz^2)\\
&+H_{\mathrm{D}6}^{1/2}H_{\mathrm{NS}5}H_{\mathrm{D}4}^{-1/2}\,d\psi^2+H_{\mathrm{D}6}^{-1/2}H_{\mathrm{NS}5}H_{\mathrm{D}4}^{1/2}\bigl(dr^2+r^2ds^2_{\tilde S^2}\bigr)\,,\\
 e^{\Phi}&=H_{\mathrm{D}6}^{-3/4}H_{\mathrm{NS}5}^{1/2}H_{\mathrm{D}4}^{-1/4}\,,\\
  H_{(3)} &= -q_{\mathrm{D}2}\text{vol}_{\text{AdS}_2}\wedge dz-q_{\mathrm{D}2} \text{vol}_{  S^2}\wedge dy+ \partial_r H_{\mathrm{NS}5}\,r^2\,d\psi \wedge\text{vol}_{ \tilde S^2}\,,\\
  F_{(2)}&=-\partial_zH_{\mathrm{D}6}dy\wedge d\psi+\partial_yH_{\mathrm{D}6}dz\wedge d\psi\,,\\
F_{(4)}&=q_{\mathrm{D}2}H_{\mathrm{D}6}\text{vol}_{\text{AdS}_2}\wedge dy \wedge d\psi-q_{\mathrm{D}2}\,H_{\mathrm{D}6} \text{vol}_{  S^2}\wedge dz\wedge d\psi+H_{\mathrm{D}6}\partial_r H_{\mathrm{D}4}\,r^2\,  dy\wedge dz\wedge \text{vol}_{ \tilde S^2}\\
&+H_{\mathrm{NS}5}\partial_y H_{\mathrm{D}4}\,r^2\,  dz\wedge d r \wedge \text{vol}_{ \tilde S^2}
+H_{\mathrm{NS}5}\partial_z H_{\mathrm{D}4}\,r^2\,  dy\wedge d r \wedge \text{vol}_{ \tilde S^2}\,, \\
F_{(6)}&=q_{\mathrm{D}2}^{4}\Bigl(\partial_y H_{\mathrm{D}4}^{-1} dy+\partial_z H_{\mathrm{D}4}^{-1}dz+\partial_r 
H_{\mathrm{D}4}^{-1} dr\Bigr) \wedge \text{vol}_{\text{AdS}_2}\wedge \text{vol}_{ S^2} \wedge d\psi\\
&+q_{\mathrm{D}2} H_{\mathrm{D}4}H_{\mathrm{NS}5} r^2 \Bigl(-\text{vol}_{\text{AdS}_2}\wedge dy+ \text{vol}_{ S^2}\wedge dz\Bigr)\wedge dr\wedge  \text{vol}_{{\tilde S}^2} \, . \\ 
\end{aligned}
\end{equation}
This defines a new class of $\ma N=4$ $\mrm{AdS}_2\times S^2\times \tilde S^2 \times \mathbb{R}^2\times S^1$ geometries fibered over a line, parametrised by $r$. The functions $H_{\text{D}4}(y,z,r)$, $H_{\text{NS}5}(r)$ and $H_{\text{D}6}(y,z)$ are solutions to the equations given in \eqref{EOM_D6NS5D4}, and describe a D4-NS5-D6 bound state localised in the subspace $\mathbb{R}_{(y,z)}^2\times \mathbb{R}_r^3$, with $\mathbb{R}_r^3$ spanned by $r$ and ${\tilde S}^2$. A solution in the class given by \eqref{brane_metric_D6D4NS5D2F1D4_nh} is thus specified by the D4-NS5-D6 charge distributions that solve these equations. The explicit study of the solutions becomes  more involved than in section \ref{IIApart}, due to the  logarithmic behaviour of $H_{\mathrm{D}6}$. Moreover, the defect interpretation found in that section in terms of semi-localised D4-NS5 branes is lost in the presence of the D6-branes, making challenging the construction of an explicit solution that asymptotes to an AdS$_5$ vacuum.

In order to proceed further with an explicit analysis of the solutions we take the  
$y$ direction inside the $\mathbb{R}_{(y,z)}^2$ plane as a circular direction, and the D6 and the D4 branes smeared along it. With this assumption the background \eqref{brane_metric_D6D4NS5D2F1D4_nh} turns out to be driven by the functions $H_{\text{D}4}(z)$, $H_{\text{NS}5}(r)$ and $H_{\text{D}6}(z)$, and the equations of motion take a much simpler form, collapsing to
\begin{equation}\label{EOM_D6NS5D4_smeared}
\partial^2_zH_{\mathrm{D}4}=0\,, \qquad \partial^2_zH_{\mathrm{D}6}=0\qquad \text{with}\qquad  H_{\mathrm{NS}5}=\frac{q_{\mathrm{NS}5}}{r}\,.
\end{equation}
Even if in this case we have not succeeded in constructing a solution with AdS$_5$ asymptotics, we show in the next subsections that it is possible to give an interpretation to a wide subclass of these solutions in terms of D2-D4$'$ baryon vertices within the 4d CFT living in D4-NS5-D6 branes.

\subsection{Quantised charges}

The most general solution to the equations of motion defined by \eqref{EOM_D6NS5D4_smeared} is that $H_{\mathrm{D}4}$ and $H_{\mathrm{D}6}$  are piecewise linear functions of $z$.  This allows one to introduce D4 and D6 source branes in the geometry. This is compatible with the quantised charges obtained from the Page fluxes, as we show below.

We start computing the F1-strings charge. The F1-strings are electrically charged with respect to the NS-NS 3-form. Their quantised charges are computed from
\begin{equation}
Q_{F1}^e=\frac{1}{(2\pi)^2}\int_{\text{AdS}_2\times I_z} H_{(3)},
\end{equation}
in units of $\alpha^\prime=g_s=1$. Regularising the volume of AdS$_2$ as $\text{Vol}_{\text{AdS}_2}=4\pi$ we find that\footnote{We use the superscript $e$ to explicitly indicate that this is an electric charge.}
\begin{equation}
Q_{F1}^e=\frac{1}{\pi} q_{\text{D}2}(z_f-z_i),
\end{equation}
where this must be computed at both ends of the $I_z$ interval. Therefore, there are $k\, q_{\text{D}2}$ F1-strings in the $z\in [0,k\pi]$ interval. We set $q_{\text{D}2}=1$ for simplicity, such that one F1-string is created as we move in $z$-intervals of length $\pi$. This is equivalent to imposing that $B_{(2)}$ lies in the fundamental region when it is integrated over AdS$_2$ (see  \cite{Lozano:2020sae}),
\begin{equation}
\frac{1}{4\pi^2}\left |\int_{\text{AdS}_2}B_{(2)}\right|\in [0,1],
\end{equation}
such that we must take
\begin{equation}
B_{(2)}^e=-(z-k\pi) \text{vol}_{\text{AdS}_2},
\end{equation}
for $z\in [k\pi,(k+1)\pi]$, for the electric part of $B_{(2)}$. The large gauge transformation parameter $k$ affects the electric components of the RR Page fluxes, $\hat{F}=F\wedge e^{-B_{(2)}}$, as
\begin{equation}
\hat{F}_{(p)}\rightarrow \hat{F}_{(p)}-k\pi F_{(p-2)}\wedge  \text{vol}_{\text{AdS}_2},
\end{equation}
in the different $[k\pi,(k+1)\pi]$ intervals.
We get, in particular
\begin{eqnarray}
&&\hat{F}_{(4)}^e=\Bigl( H_{\mathrm{D}6}-(z-k\pi)\partial_z H_{\mathrm{D}6}\Bigr) \text{vol}_{\text{AdS}_2}\wedge dy\wedge d\psi\nonumber\\
&&\hat{F}_{(6)}^e=-H_{\mathrm{NS}5}r^2 \Bigl( H_{\mathrm{D}4}-(z-k\pi)\partial_z H_{\mathrm{D}4}\Bigr) \text{vol}_{\text{AdS}_2}\wedge dr\wedge  \text{vol}_{{\tilde S}^2},\label{Felectric}
\end{eqnarray}
for $z\in [k\pi,(k+1)\pi]$.
These electric fluxes give rise to D2 and D4$'$ electric charges, respectively, that we compute through
\begin{equation}
Q^e_{\mathrm{D}p}=\frac{1}{(2\pi)^{p+1}}\int_{\text{AdS}_2\times \Sigma_p}\hat{F}_{(p+2)}^e,
\end{equation}
in units of $\alpha^\prime=g_s=1$. 

For the D4 and D6 branes  it will be more convenient to compute their magnetic charges, associated to the magnetic components of $\hat{F}_{(4)}$ and $\hat{F}_{(2)}$ given by
\begin{eqnarray}
&&\hat{F}_{(4)}^m=-H_{\mathrm{NS}5}\partial_z H_{\mathrm{D}4}r^2 dy\wedge dr\wedge \text{vol}_{{\tilde S}^2}\nonumber\\
&&\hat{F}_{(2)}^m=-\partial_z H_{\mathrm{D}6} dy\wedge d\psi\label{Fmagnetic}\, .
\end{eqnarray}
In the presence of sources the Bianchi identities in equation \eqref{EOM_D6NS5D4_smeared} are modified such that
\begin{eqnarray}
&&d\hat{F}_{(4)}=\partial_z^2 H_{\mathrm{D}4} dz \wedge dy \wedge dr \wedge \text{vol}_{{\tilde S}^2}\\
&&d\hat{F}_{(2)}=\partial_z^2 H_{\mathrm{D}6} dz \wedge dy\wedge d\psi.
\end{eqnarray}
Therefore, the D4 and D6 branes provide sources localised in the $z$ direction. They are thus flavour branes. They span, respectively, the 
$\text{AdS}_2\times S^2\times S^1_\psi$ and $\text{AdS}_2\times S^2\times I_r\times {\tilde S}^2$ subspaces of the solution (see below).

We need to specify now the  linear functions $H_{\mathrm{D}4}$ and $H_{\mathrm{D}6}$. We take both functions to be piecewise linear in the different $z\in [k\pi,(k+1)\pi]$ intervals, with the space starting and ending at $z=0$ and $z=\pi (P+1)$, where both
$H_{\mathrm{D}4}$ and $H_{\mathrm{D}6}$ are taken to vanish. This parallels the analysis done in \cite{Lozano:2020txg,Lozano:2020sae} for the AdS$_2$ solutions constructed therein, based, in turn, in the field theory interpretation of the AdS$_3$ solutions constructed in \cite{Lozano:2019emq}, carried out in \cite{Lozano:2019jza,Lozano:2019zvg}\footnote{Reference \cite{Lozano:2021xxs} is a recent review article which summarises these developments.}.
In this way the singularity structure at both ends of the space,
\begin{equation}
ds^2\sim x^{-1}( ds^2_{\text{AdS}_2}+ds^2_{S^2})+x(dy^2+dx^2)+d\psi^2+dr^2+r^2ds^2_{{\tilde S}^2}, \qquad e^\Phi\sim x^{-1},
\end{equation}
where $x= z$ close to $z=0$ and $x=\pi(P+1)-z$ close to $z=\pi (P+1)$, corresponds to a superposition of D4-branes wrapped on $\text{AdS}_2\times S^2\times S^1_\psi$ and smeared on $(y, r, {\tilde S}^2)$, and D6 branes wrapped on $\text{AdS}_2\times S^2\times I_r\times {\tilde S}^2$ and smeared on $(\psi,y)$\footnote{Note that the same behaviour is obtained from a superposition of O4 and O6 orientifold fixed planes.}. The $H_{\mathrm{D}4}$ and $H_{\mathrm{D}6}$ functions then read
\begin{equation} \label{profileh4}
H_{\mathrm{D}4}(z)=\left\{ \begin{array}{ccrcl}
                       \frac{\beta_0 }{\pi}
                       z & 0\leq z\leq \pi \\
                                     \alpha_k\! +\! \frac{\beta_k}{\pi}(z-\pi k) &~~ \pi k\leq z \leq \pi(k+1),\;\; k=1,..,P-1\\
                      \alpha_P-  \frac{\alpha_P}{\pi}(z-\pi P) & \pi P\leq z \leq \pi(P+1),
                                             \end{array}
\right.
\end{equation}
 \begin{equation} \label{profileh6}
H_{\mathrm{D}6}(z)
                    =\left\{ \begin{array}{ccrcl}
                       \frac{\nu_0 }{\pi}
                       z & 0\leq z\leq \pi \\
                        \mu_k+ \frac{\nu_k}{\pi}(z-\pi k) &~~ \pi k\leq z \leq \pi(k+1),\;\;\;\; k=1,....,P-1\\
                      \mu_P-  \frac{\mu_P}{\pi}(z-\pi P) & \pi P\leq z \leq \pi(P+1),
                                             \end{array}
\right.
\end{equation}
where $\alpha_k, \beta_k, \mu_k, \nu_k$ are integration constants, which must satisfy (see  \cite{Lozano:2020sae})
\begin{equation}
\alpha_k=\sum_{j=0}^{k-1}\beta_j,\qquad \mu_k=\sum_{j=0}^{k-1}\nu_j, \label{defalphamu}
\end{equation}
for $k=1,\dots, P$,
for continuity of the metric and dilaton. In turn, the fluxes can have discontinuities associated to the presence of branes. 

Note that in order to find well defined quantised charges from the electric and magnetic fluxes in \eqref{Felectric} and \eqref{Fmagnetic} we need to globally define the $r$-direction as well. We do this by taking the $(r,{\tilde S}^2)$ space to span a 3-torus $T^3$. We then find  the quantised charges
\begin{eqnarray}
&&Q_{D2}^{e(k)}=\mu_k, \qquad Q_{D4'}^{e(k)}=\alpha_k, \qquad Q_{F1}^{e(k)}=1, \label{quantised1}\\
&&Q_{D4}^{m(k)}=\beta_k, \qquad Q_{D6}^{m(k)}=\nu_k, \label{quantised2}
\end{eqnarray}
in the different $z\in [k\pi,(k+1)\pi]$ intervals.
Here we have also used that $y\in [0,\pi]$\footnote{The reason for this particular periodicity will become clear in the conclusions, when we discuss the relation between these geometries and the double analytical continuation of the $\text{AdS}_3\times S^2$ geometries studied in  \cite{Faedo:2020lyw}.}. The equations \eqref{quantised1}, \eqref{quantised2} show that the integration constants $\alpha_k, \beta_k, \mu_k, \nu_k$ must be integer numbers. Moreover, we see that the D2 and D4$'$ charges at each $z\in [k\pi,(k+1)\pi]$ interval are equal to the sum of the D6 and D4 charges, respectively, in the previous $[0,k\pi]$ intervals. We will find an explanation for this fact when we give the field theory interpretation to these solutions in the next subsections. The number of D4 and D6 source branes present at each interval is given by
\begin{equation}
F_k=\beta_{k-1}-\beta_k, \qquad {\tilde F}_k=\nu_{k-1}-\nu_k, \label{quantised3}
\end{equation}
consistently with the derivatives
\begin{equation}
\partial_z^2 H_{\mathrm{D}4}=\frac{1}{\pi}\sum_{k=1}^P (\beta_{k-1}-\beta_k)\delta(z-\pi k), \qquad
\partial_z^2 H_{\mathrm{D}6}=\frac{1}{\pi}\sum_{k=1}^P (\nu_{k-1}-\nu_k)\delta(z-\pi k), 
\end{equation}
that follow from \eqref{profileh4} and \eqref{profileh6}.

Finally, we would like to briefly discuss the role played by the NS5 and NS5$'$ branes in the brane set-up. As we have discussed, the D2 and D4$'$ branes play the role of colour branes of the configuration. The D2 branes are wrapped on the two circular directions $(y,\psi)$. They are stretched in the $y$ direction between NS5$'$ branes, that are located at $y=0,\pi$ and are periodically identified, and between two NS5 branes\footnote{We take $q_{\text{NS}5}=1$ for simplicity.} in the $\psi$ direction, located at $\psi=0,2\pi$ and periodically identified. The field theory that lives in them is therefore one dimensional at low energies. In turn, there are D4$'$ branes wrapped on $y$ and on the $T^3$. They are stretched in the $y$ direction between NS5$'$ branes. The field theory that lives in them is therefore also one dimensional at low energies. Our conjecture is that the $\ma N=4$ quantum mechanics that lives in the D2-D4$'$ branes flows to a super conformal quantum mechanics in the IR that is dual to the backgrounds defined by the functions \eqref{profileh4}, \eqref{profileh6}. We turn to this analysis in subsection \ref{dualQM}, after discussing the holographic central charge in the next subsection.

\subsection{Holographic central charge}

Defining a central charge for a one dimensional CFT is known to be a subtle issue, see for instance \cite{Hartman:2008dq,Alishahiha:2008tv,Cadoni:1999ja}. There is however a notion of holographic ``central charge'' that can be computed from the Brown-Henneaux formula,
\begin{equation}
c_{\text{holo}}=\frac{3V_{int}}{4\pi G_N},
\end{equation}
where $G_N$ is the ten dimensional Newton's constant, $G_N=8\pi^6$, and $V_{int}$ is the volume of the internal space. In the presence of a dilaton term the internal volume must be modified, following the usual procedure (see \cite{Klebanov:2007ws,Macpherson:2014eza}). For our class of backgrounds we find
\begin{equation} 
c_{\text{holo}}= \frac{3}{4\pi G_N} \int d\vec\theta \sqrt{e^{-4\Phi}\det(g_{ij})} = \frac{6}{\pi}   \int_0^{\pi(P+1)} H_{\mathrm{D}4}H_{\mathrm{D}6} \, dz,
\end{equation}
where $g_{ij}$ is the metric of the inner space and $\vec\theta$ are coordinates defined over it. Substituting here the expressions for $H_{\mathrm{D}4}$ and $H_{\mathrm{D}6}$ given by \eqref{profileh4} and \eqref{profileh6} we arrive at the final expression,
\begin{equation} 
c_{\text{holo}} = 6 \sum_{k=0}^{P} \left(\alpha_k \mu_k + \frac13 \beta_k \nu_k 
+ \frac12( \alpha_k\nu_k + \beta_k \mu_k) \right), \label{holocc}
\end{equation}
that we will compare with the ``central charge'' of the superconformal quantum mechanics in subsection \ref{Qmcc}.

\subsection{Dual quiver quantum mechanics}\label{dualQM}

In this subsection we propose quiver quantum mechanics supported by our solutions. The dynamics of the quivers is described in terms of the matter fields associated to the open strings that connect the different branes.  We will follow closely the detailed description of the matter fields given in  \cite{Lozano:2020sae} (see Appendix B therein), since our brane system is related by two T-dualities (along the $y$ and $\psi$ directions) to the D0-D4-F1-D4$'$-D8 brane intersections studied there. As in that reference we will use 2d $\ma N=(0,4)$ notation for the 1d $\ma N=4$ matter fields.

\begin{figure}[h!]
    \centering
    {{\includegraphics[width=10cm]{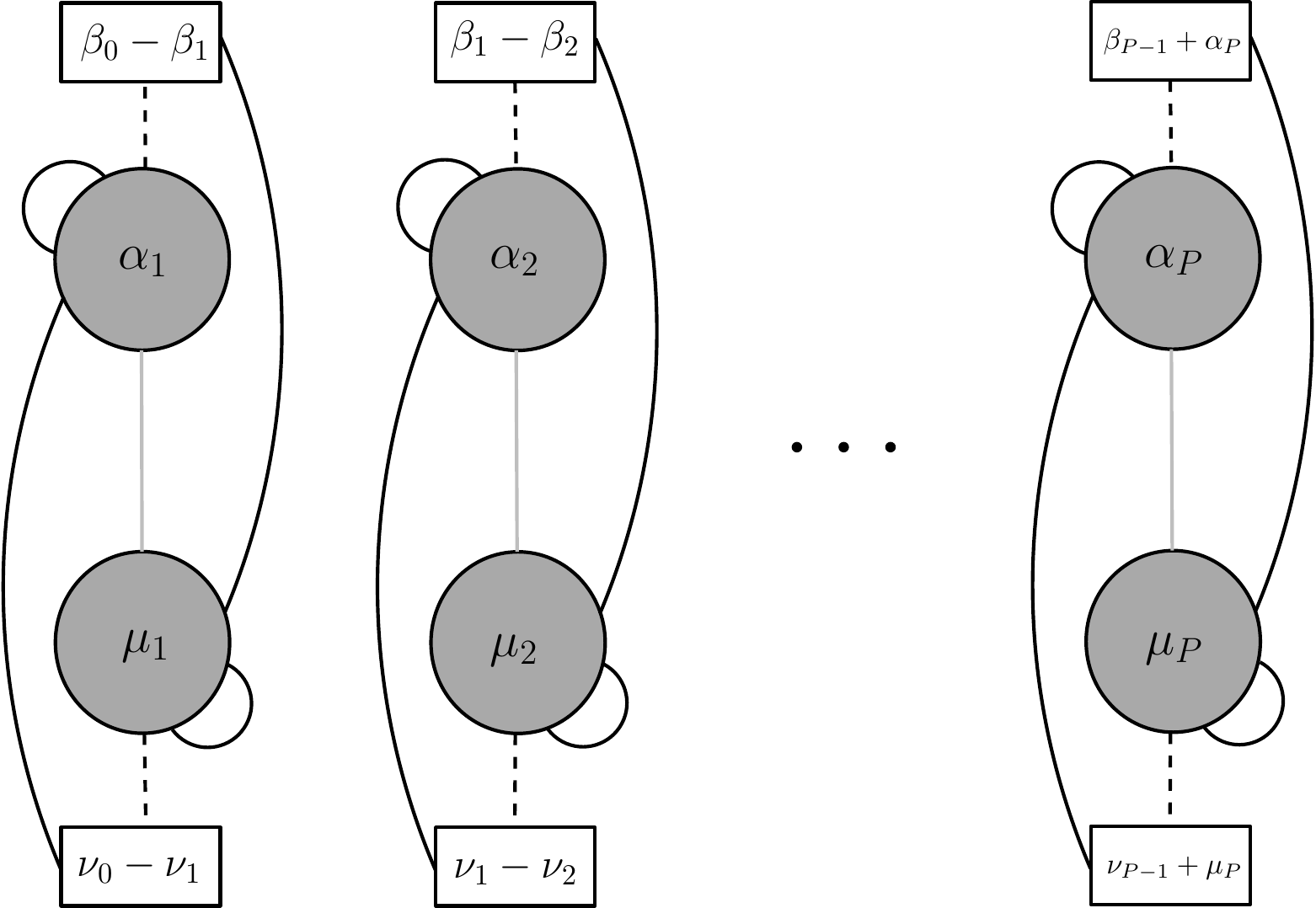} }}%
    \caption{Disconnected quivers describing the SCQMs dual to our solutions.}
\label{quivers}
\end{figure}

As all D-branes are localised in the $z$-direction, strings stretched between branes in adjacent $[\pi k,\pi (k+1)]$ intervals are long, and describe massive states. Therefore, they will not show in the quiver quantum mechanics. We will discuss their role in the field theory in subsection \ref{baryonvertex}. Therefore, the full Hilbert space of the system is given by the sum of the individual Hilbert spaces of the D2-D4$'$-D4-D6 branes in each $[\pi k,\pi (k+1)]$ interval, whose degrees of freedom we summarise next:

\begin{itemize}
\item D2-D2: Given that the D2 branes are wrapped on the $y$ and $\psi$ directions they are effectively point like. They contribute with a (4,4) vector and a (4,4) hypermultiplet in the adjoint. 
\item D4$'$-D4$'$: Given that the D4$'$ branes are wrapped on $y$ and on the $T^3$ they are also effectively point like. They contribute as well with a (4,4) vector and a (4,4) hypermultiplet in the adjoint. 
\item D2-D4$'$: The D2-D4$'$ subsystem is T-dual to the D0-D4 system in  \cite{Lozano:2020sae}. They contribute with a (4,4) hypermultiplet in the bifundamental representation of the two gauge groups.
\item D2-D4: This subsystem is T-dual to the D0-D4$'$ system in \cite{Lozano:2020sae}. They contribute with a (4,4) bifundamental twisted hypermultiplet.
\item D2-D6: This is T-dual to D0-D8. They contribute with a (0,2) bifundamental Fermi multiplet.
\item D4$'$-D4: This is T-dual to D4-D4$'$. They contribute with a (0,2) bifundamental Fermi multiplet.
\item D4$'$-D6: This is T-dual to D4-D8. They contribute with a (4,4) bifundamental twisted hypermultiplet.
\end{itemize}
Putting all this together, and using that the ranks of the gauge and flavour groups associated to the D2-D4$'$-D4-D6 branes in a given $[\pi k,\pi (k+1)]$ interval are given by $\mu_k$, $\alpha_k$,  $\beta_{k-1}-\beta_k$ and $\nu_{k-1}-\nu_k$, respectively (see equations \eqref{quantised1}, \eqref{quantised2}, \eqref{quantised3}),  
we get the field content depicted in Figure \ref{quivers}. In this figure circles represent (4,4) vector multiplets, black lines (4,4) twisted hypermultiplets, grey lines (4,4) hypermultiplets and dashed lines (0,2) Fermi multiplets.  Note that this is the same quiver quantum mechanics discussed in \cite{Lozano:2020sae} (see section 3.3 therein), now associated to a different brane system. 
As in that paper, the quantum mechanics will find an interpretation as describing Wilson lines (more specifically,  baryon vertices) within a higher dimensional QFT, once we introduce the massive F1-strings stretched between the branes in the $z$-direction. We turn to this description in subsection \ref{baryonvertex}. Before doing that we briefly address the computation of the quantum mechanical central charge.

\subsection{Quantum mechanical central charge}\label{Qmcc}

In this subsection we address the computation of the central charge, following closely \cite{Lozano:2020sae}. The usual caveats involved in  the definition of a superconformal quantum mechanical central charge are present in our current set-up. The central charge should then be interpreted as counting the degeneracy of ground states of the system. 

We will follow the proposal in \cite{Lozano:2020txg}.
In that reference the central charge of a 1d CFT arising as a chiral half of a 2d (0,4) CFT was computed using the same expression that allows to count the degrees of freedom of the original 2d CFT, given by \cite{Putrov:2015jpa},
\begin{equation}
c=6\, (n_{hyp}-n_{vec}) \label{formulaputrov}.
\end{equation}
It was argued that since $\ma N=4$ multiplets arise upon reduction of 2d (0,4) multiplets, this expression can be used to account for the degrees of freedom of the 1d CFT through direct counting of its $\ma N=4$ hypermultiplets and vector multiplets.
More interestingly for our purposes, even if this formula is valid, by construction, for 1d CFTs with a 2d (0,4) origin, it was proposed in \cite{Lozano:2020sae} that it could also be used to count the number of ground states of purely 1d CFTs. In that reference it was noted that \eqref{formulaputrov} extends to more general SCQM the formula derived in \cite{Denef:2002ru,Ohta:2014ria,Cordova:2014oxa} for the dimension of the Higgs branch 
of so-called Kronecker quivers, built out of 1d gauge groups connected by bifundamentals, given by
\begin{equation}
{\cal M}=\sum_{v,w}N_v N_w -\sum_v N_v^2+1. \label{Higgs}
\end{equation}
In this expression $N_w$ stands for the ranks of the colour groups adjacent to a given colour group of rank $N_v$. It is straightforward to check that for these quivers  the dimension of the Higgs branch computed from \eqref{Higgs} agrees with the central charge computed using \eqref{formulaputrov}, up to a factor of 1\footnote{This factor of 1 is irrelevant in the holographic limit, but we are lacking a precise understanding of the origin of this discrepancy.} and the global normalisation.
The quantum mechanical central charge computed from \eqref{formulaputrov} was shown to agree to leading order with the corresponding holographic expression in a number of examples \cite{Lozano:2020sae}.

We will thus use \eqref{formulaputrov} to compute the central charge of our quiver quantum mechanics. In this computation, as remarked in \cite{CLPV}, $n_{hyp}$ counts the number of ordinary (as opposed to twisted) $\ma N=4$ hypermultiplets, since the U$(1)_R$-charge of the fermions in twisted hypermultiplets vanishes. For the quivers depicted in Figure \ref{quivers} we find that
\begin{equation}
n_{hyp}=\sum_{k=1}^P (\alpha_k \mu_k + \alpha_k^2+\mu_k^2), \qquad n_{vec}=\sum_{k=1}^P (\alpha_k^2+\mu_k^2),
\end{equation}
and therefore,
\begin{equation}
c=6\sum_{k=1}^P \alpha_k \mu_k,
\end{equation}
identically.
Keeping in mind the definitions of $\alpha_k$, $\mu_k$, given by \eqref{defalphamu}, we find that this expression agrees in the large number of nodes limit with the holographic expression, given by \eqref{holocc}. Moreover, the agreement is exact in the absence of any of the two types of flavour branes. It would be interesting to have a more precise understanding of this exact agreement.

\subsection{Baryon vertex interpretation}\label{baryonvertex}

In this subsection we turn our attention to the interpretation of the massive F1-strings. The discussion will again follow very closely the field theory interpretation given to the AdS$_2$ solutions constructed in \cite{Lozano:2020sae}. The key point is to realise that the orientation between the D4 and the D4' branes, and between the D6 and the D2 branes in the brane set-up is the one that allows to create F1-strings stretched between the D4 and the D4$'$ branes and between the D6 and the D2 branes. 
These strings have as their lowest energy excitation a fermionic field, which upon integration leads to a Wilson loop. 

In \cite{Yamaguchi:2006tq,Gomis:2006im} it was shown that a half-BPS Wilson loop in a U($N$) antisymmetric representation of 4d $\ma N=4$ SYM  can be described by an array of $M$ D5-branes with fundamental strings dissolved in their worldvolumes. This is the realisation in the near horizon limit of a configuration of $M$ stacks of D5-branes separated a distance $L$ from $N$ D3-branes, with $(m_1, m_2, \dots m_M)$ F1-strings stretched between the stacks. The brane set-up is depicted in Table \ref{Wilsonloop}.
\begin{table}[http!]
\renewcommand{\arraystretch}{1}
\begin{center}
\scalebox{1}[1]{
\begin{tabular}{cc ||cc c c  c  c c c c}
 branes & $t$ & $x^1$ & $x^2$ & $x^3$ & $x^4$ & $x^5$ & $x^6$ & $x^7$ & $x^8$ & $x^9$ \\
\hline \hline
$\mrm{D}3$ & $\times$ & $\times$ & $\times$ & $\times$ & $-$ & $-$ & $-$ & $-$ & $-$ & $-$ \\
$\mrm{D}5$ & $\times$ & $-$ & $-$ & $-$ & $-$ & $\times$ & $\times$ & $\times$ & $\times$ & $\times$ \\
$\mrm{F}1$ & $\times$ & $-$ & $-$ & $-$ & $\times$ & $-$ & $-$ & $-$ & $-$ & $-$ \\
\end{tabular}
}
\caption{Brane set-up associated to the D3-D5-F1 brane configuration that describes Wilson loops in antisymmetric representations of 4d U($N$) $\ma N=4$ SYM. } \label{Wilsonloop}
\end{center}
\end{table}
As one can easily check this is precisely the relative orientation between the D4, the F1 and the D4$'$ branes in Table 
 \ref{Table:branesinmasslessIIA1D6} and the D6, the F1 and the D2 branes. 
 
\begin{figure}[h!]
    \centering
    {{\includegraphics[width=10cm]{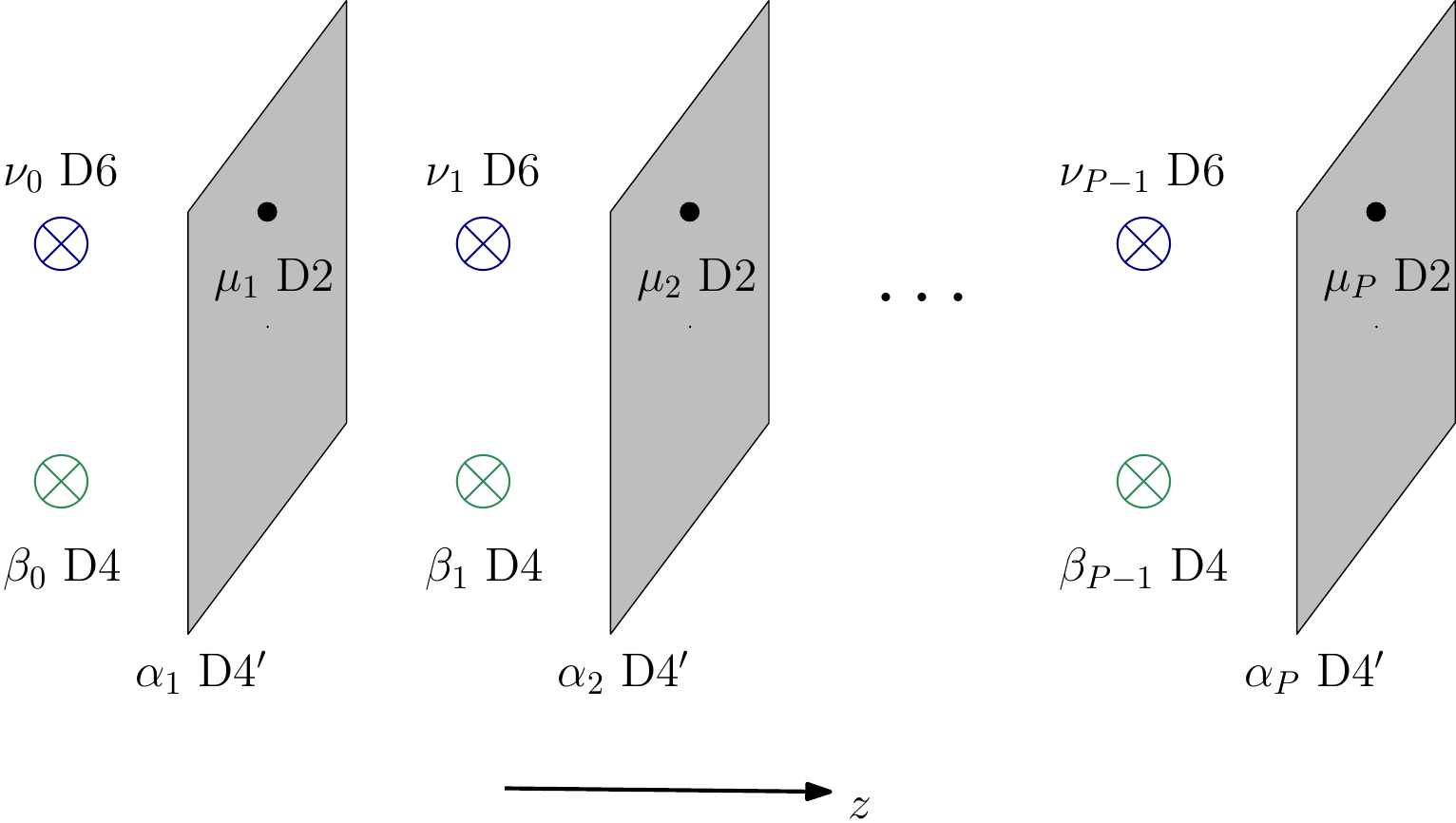} }}%
    \caption{Hanany-Witten brane set-up associated to the quantised charges of the solutions.}
\label{HW1}
\end{figure}

 Indeed, the couplings that describe Wilson lines in the worldvolumes of the D4$'$ and D2 colour branes are, respectively,
 \begin{equation}
 S_{\text{D}4'}=T_4\int \hat{F}_{(4)} \wedge A_t, \qquad S_{\text{D}2}=T_2 \int \hat{F}_{(2)} \wedge A_t.
 \end{equation}
In the first expression the D4$'$ branes are wrapped on $y$ and the $T^3$, therefore they capture the $\hat{F}_{(4)}^m$ magnetic flux given in \eqref{Fmagnetic}. In turn, the D2 branes are wrapped on $y$ and $\psi$, so they capture the $\hat{F}_{(2)}^m$ magnetic flux. Substituting these fluxes in the $[\pi k, \pi (k+1)]$ $z$-interval we arrive at
\begin{equation}
S_{\text{D}4'}=\beta_k T_{F1}\int dt A_t, \qquad S_{\text{D}2}=\nu_k T_{F1}\int dt A_t.
\end{equation}
These expressions describe, respectively, $\beta_k$ and $\nu_k$ Wilson lines. If we add now the contributions of the F1-strings stretched between the D4$'$ branes in the $k$ interval and the D4 branes in all previous intervals, and the same for the D2 branes and the D6 branes, as depicted in Figure \ref{HW1},
we find Wilson lines in the $(\beta_0,\beta_1,\dots, \beta_{k-1})$ and $(\nu_0,\nu_1,\dots, \nu_{k-1})$ antisymmetric representations of the U($\alpha_k$) and U($\mu_k$) gauge groups. This is precisely the realisation of the baryon vertices associated to these gauge groups. 

\begin{figure}[h!]
    \centering
    {{\includegraphics[width=11cm]{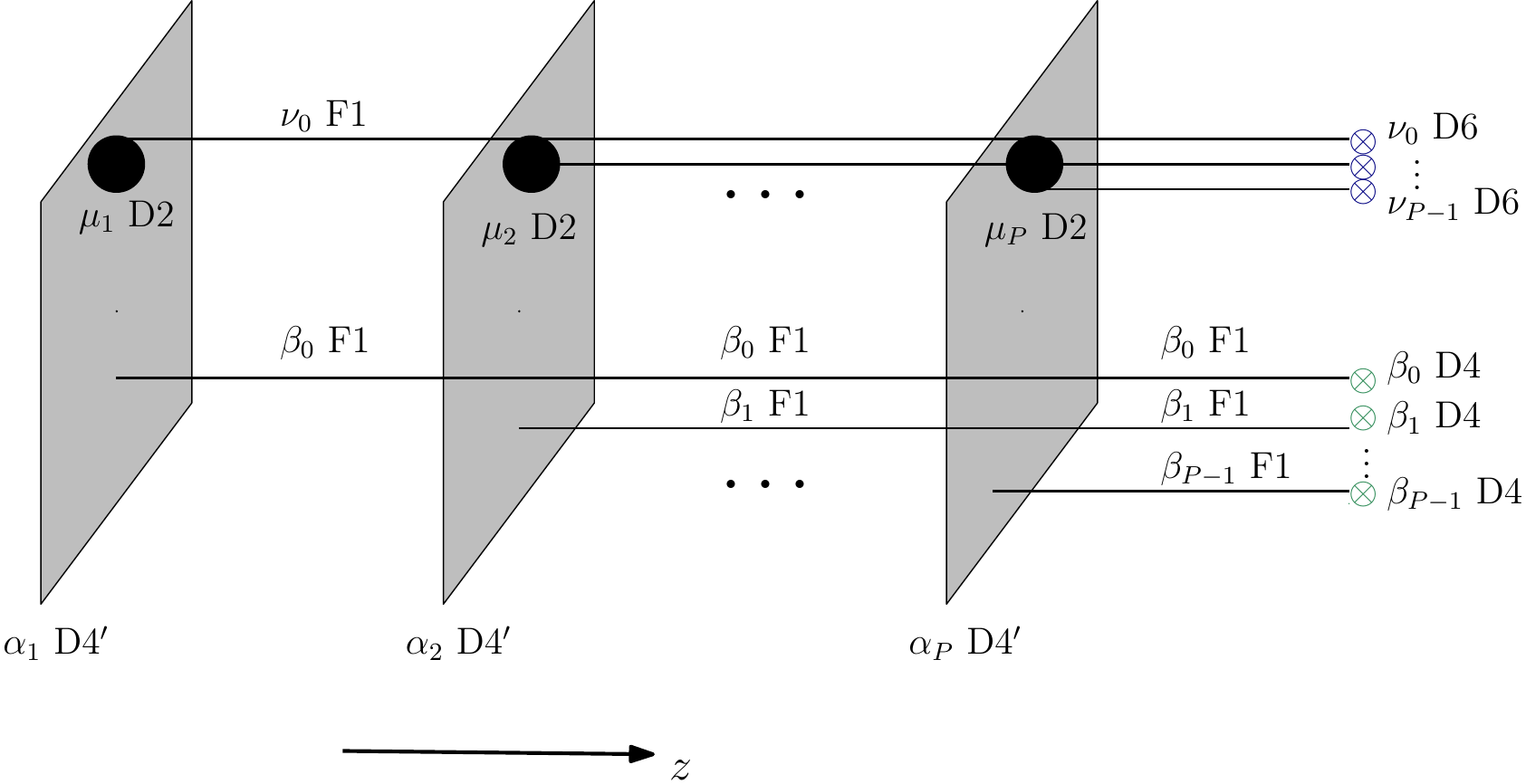} }}%
    \caption{Hanany-Witten brane set-up equivalent to the brane configuration in Figure \ref{HW1}.}
\label{HW2}
\end{figure}

Indeed, the brane set-up depicted in Figure \ref{HW1}, 
can be related after the combination of a T-duality, an S-duality, successive  Hanany-Witten moves and a further T-duality to the brane set-up depicted in Figure \ref{HW2}.
This relation is carefully explained in \cite{Lozano:2020sae}. For the D4$'$-F1-D4 brane subsystem it follows directly from the analysis of the  D4$'$-F1-D4 brane  system in \cite{Lozano:2020sae}\footnote{Note that the D4$'$ and the D4 branes are interchanged in that reference.}, while for the D2-F1-D6 subsystem it follows from the analysis of the D0-F1-D8 subsystem therein after two T-dualities. The reader can find more details about this description in that reference.

Our previous description is consistent with an interpretation of the AdS$_2$ solutions given by \eqref{brane_metric_D6D4NS5D2F1D4_nh}, with the profiles specified in \eqref{profileh4}, \eqref{profileh6},  as describing backreacted baryon vertices within the 4d $\ma N=2$ CFT living in the D4-NS5-D6 branes. In this interpretation the SCQM arises in the very low energy limit of a system of D4-NS5-D6 branes in which one dimensional defects are introduced. The one dimensional defects consist on D4$'$ baryon vertices, connected to the D4-branes with F1-strings, and D2-brane baryon vertices, connected to the D6 by F1-strings. In the IR the gauge symmetry on the D4 branes becomes global, turning them from colour to flavour branes. In turn, the D4$'$ and the D2 defect branes  become the new colour branes of the backreacted geometry. This interpretation goes in parallel with the proposed defect interpretation for the classes of AdS$_2$ solutions found in \cite{Lozano:2020sae} and \cite{Lozano:2021rmk}. Interestingly, for the first class of geometries the AdS$_6$ solution of Brandhuber-Oz \cite{Brandhuber:1999np} was shown to arise locally far away from the defect \cite{Dibitetto:2018gtk}. In our case we should be able to find the AdS$_5$ geometry dual to the D4-NS5-D6 brane intersection far away from the defect. This  is currently under investigation.

\section{Conclusions}\label{conclusions}

In this paper we have constructed new families of $\text{AdS}_2\times S^2\times S^2\times \mathbb{R}^2\times S^1\times I$ solutions to Type II supergravities preserving four Poincar\'e supersymmetries. Starting with Type IIA supergravity we have constructed solutions corresponding to the intersection of D2-F1-D4$'$-NS5$'$ branes ending on a bound state of D4-NS5 branes. Taking the near horizon limit a vast class of solutions of the type specified above has been shown to emerge, determined by the charge distribution of the D4-NS5 system. Choosing a particular semi-localised profile for the D4-NS5 branes we have then shown that two interesting regimes arise. The first regime allows one to approach the D2-F1-D4$'$-NS5$'$ defect branes. In this limit the defect branes are resolved into the fully backreacted $\text{AdS}_2\times S^2\times S^2\times \mathbb R^2\times S^1\times I$ geometry. The second regime becomes manifest in a new system of coordinates, and allows one to move away from the defect branes. In this regime an AdS$_5$ geometry arises asymptotically, which corresponds to the near horizon geometry of the D4-NS5 branes. We have shown that the particular AdS$_5$ vacuum that emerges in this limit is the T-dual of the $\text{AdS}_5\times S^5/\mathbb{Z}_n$ solution to Type IIB supergravity, holographically dual to 4d $\ma N=4$ SYM modded by $\mathbb{Z}_n$. This has allowed us to interpret this class of solutions as dual to line defect CFTs within 4d $\ma N=4$ SYM modded by $\mathbb{Z}_n$.

Inspired by our findings in Type IIA we have performed a similar analysis in Type IIB supergravity. In this case we have constructed new families of  
$\text{AdS}_2\times S^2\times S^2\times \mathbb{R}^2\times S^1\times I$ solutions with the same supersymmetries, now emerging in the near horizon limit of D1-F1-D5-NS5 branes ending on D3 branes probing an ALE singularity. In the second limit described above the $\text{AdS}_5\times S^5/\mathbb{Z}_n$ solution of Type IIB supergravity emerges asymptotically. The solutions thus admit as well a line defect interpretation, this time in terms of D1-F1-D5-NS5 defect branes.

Returning to Type IIA, we have extended our class of solutions to include D6 branes. In this case the defect interpretation in terms of semi-localised D4-NS5 branes is lost, and we have not succeeded in constructing explicit solutions that asymptote to an AdS$_5$ vacuum. Instead, we have turned into a thorough study of the 1d CFTs dual to these general  solutions, taking a simplified ansatz.
These CFTs arise in the IR limit of explicit quiver quantum mechanics that we have constructed.  Remarkably, these are the same quiver quantum mechanics that flow in the IR to the 1d CFTs  studied in \cite{Lozano:2020sae}, dual to the $\text{AdS}_2\times S^3\times \text{CY}_2\times I$ solutions to massive Type IIA with $\ma N=4$ supersymmetries studied therein. This happens because when the branes are smeared on $y$ (the simplified assumption taken in our construction) our brane system is related by two T-dualities to the brane system discussed in \cite{Lozano:2020sae}, consisting on a D0-F1-D4-D4$'$-D8 intersection. The 1d dual CFT associated to this class of solutions was interpreted in terms of baryon vertices within 5d fixed point theories living in D4-D8 intersections, and an asymptotically locally AdS$_6$ geometry was shown to arise for certain solutions. Our findings in this paper show that the same quiver quantum mechanics describe in the UV line defect CFTs associated to baryon vertices within 4d $\ma N=2$ SCFTs. In this last case we are lacking however a defect completion within an AdS$_5$ vacuum in Type IIA. 

Still, the smearing in $y$ allows to make connection with the class of AdS$_3\times S^2\times S^2 \times S^1\times \Sigma_2$ solutions to Type IIB supergravity studied in  \cite{Faedo:2020lyw}, for which an interpretation as surface defects within $\mrm{AdS}_6\times S^2 \times \Sigma_2$ Type IIB vacua was found. Indeed, T-dualising our solutions along the $y$ direction a new class of AdS$_2\times S^3\times S^2 \times S^1 \times \Sigma_2$ backgrounds in Type IIB supergravity with the same number of supersymmetries is produced, associated to D1-F1-D3-D5-NS5-D7 intersections. These backgrounds are related through a double analytic continuation to the solutions studied in  \cite{Faedo:2020lyw}, and it is expected that the same 
$\mrm{AdS}_6\times S^2 \times \Sigma_2$ vacua will arise asymptotically. This would allow us to interpret our solutions as dual to line defect CFTs within the 5d fixed point theories living in D5-NS5-D7 intersections. This is currently under investigation 
 \cite{LozanoPrep}.

\section*{Acknowledgements}

We would like to thank Chris Couzens and Stefan Vandoren for very useful discussions. The authors are partially supported by the Spanish government grant PGC2018-096894-B-100.

\appendix

\section{M-Theory picture}
\label{appMtheory}
In this appendix we provide the M-theory interpretation of the defect solution studied in section \ref{IIApart}.
We start discussing the brane interpretation of warped $\ma N=2$ AdS$_5$ vacua in M-theory \cite{Papadopoulos:1996uq,Gauntlett:1996pb,Tseytlin:1996bh,Youm:1999ti,Fayyazuddin:1999zu,Loewy:1999mn,Alishahiha:1999ds,Oz:1999qd}. The strong coupling limit of the D4-NS5 system discussed in section \ref{AdS5semilocalD4NS5} has a simple interpretation in 11d as the bound state of M5 branes depicted in Table \ref{Table:M5M5system}.
\begin{table}[http!]
\renewcommand{\arraystretch}{1}
\begin{center}
\scalebox{1}[1]{
\begin{tabular}{c c cc  c|| c c c  c |c c c}
 branes & $t$ & $x^1$ & $x^2$ & $x^3$ & $y$ & $z$ & $\psi$ & $\chi$ & $r$ & $\theta^1$ & $\theta^2$ \\
\hline \hline
$\mrm{M}5_1$ & $\times$ & $\times$ & $\times$ & $\times$ & $-$ & $-$ & $\times$ &$\times$ & $-$ & $-$ & $-$ \\
$\mrm{M}5_2$ & $\times$ & $\times$ & $\times$ & $\times$ & $\times$ & $\times$ & $-$ & $-$ & $-$ & $-$ & $-$ \\
\end{tabular}
}
\caption{M-theory picture underlying warped $\ma N=2$ AdS$_5$ vacua. This intersection preserves 8 real supercharges and supports a 4d $\ma N=2$ SCFT. $\chi$ parametrises the M-theory circle.} \label{Table:M5M5system}
\end{center}
\end{table}
This system can be obtained geometrically by uplifting the D4-NS5 bound state of Table \ref{Table:D4NS5system}. The D4 branes become M5$_1$ branes localised within the space $(\mathbb{R}^2_{(y,z)},\mathbb{R}^3_r )$, and the NS5 branes M5$_2$ branes completely delocalised on $\psi$ and $\chi$. Note that in M-theory $\psi$ can naturally parametrise an interval instead of a circle, such that an $\mathbb{R}^2_{(\psi, \chi)}$ plane emerges in the strong coupling limit in which the M-theory circle decompactifies\footnote{In this case the circle where the M5$_1$ branes are wrapped appears by rewriting $\mathbb{R}^2_{(\psi, \chi)}$ in polar coordinates.}.
The metric and fluxes of such a bound state are given by
\begin{equation}
\label{brane_metric_M5M5}
\begin{split}
d s_{11}^2 &= H_{\mathrm{M}5_1}^{-1/3}H_{\mathrm{M}5_2}^{-1/3}  ds^2_{\mathbb{R}^{1,3}} + H_{\mathrm{M}5_1}^{2/3}H_{\mathrm{M}5_2}^{-1/3}(dy^2+dz^2)\\
&+H_{\mathrm{M}5_1}^{-1/3}H_{\mathrm{M}5_2}^{2/3}\,(d\psi^2+d\chi^2)+H_{\mathrm{M}5_1}^{2/3}H_{\mathrm{M}5_2}^{2/3} \bigl(dr^2+r^2ds^2_{\tilde S^2}\bigr)\,,\\
G_{(4)}&=\partial_r H_{\mathrm{M}5_1}\,r^2\,  dy\wedge dz \wedge \text{vol}_{ \tilde S^2}
+H_{\mathrm{M}5_2}\partial_y H_{\mathrm{M}5_1}\,r^2\,  dz\wedge d r \wedge \text{vol}_{ \tilde S^2}\\
&-H_{\mathrm{M}5_2}\partial_z H_{\mathrm{M}5_1}r^2\, dy\wedge d r \wedge \text{vol}_{ \tilde S^2}+\partial_r H_{\mathrm{M}5_2}r^2d\psi\wedge d\chi \wedge\text{vol}_{ \tilde S^2}\,,\\
\end{split}
\end{equation}
where the harmonic functions are such that $H_{\mathrm{M}5_1}(y,z,r)= H_{\mathrm{D}4}(y,z,r)$ and $H_{\mathrm{M}5_2}(r)= H_{\mathrm{NS}5}(r)$. As for the D4-NS5 system one can construct a semi-localised solution with harmonic functions  \cite{Youm:1999ti,Loewy:1999mn}
\begin{equation}\label{semilocalizedM51M52}
 H_{\mathrm{M}5_1}=1+\frac{q_{\mathrm{M}5_1}}{(y^2+z^2+4 q_{\mathrm{M}5_2}r)^2}\qquad \text{and}\qquad H_{\mathrm{M}5_2}=\frac{q_{\mathrm{M}5_2}}{r}\,.
\end{equation}
Making the change of coordinates \eqref{AdS5coord}, which in M-theory notation takes the form \cite{Oz:1999qd},
\begin{equation}\label{changeMtheory}
   y= \mu \sin \alpha\cos\phi\,,\qquad z= \mu \sin \alpha\sin\phi\qquad  \text{and}\qquad r= 4^{-1}\,q_{\mathrm{M}5_2}^{-1}\, \mu^2\cos^2\alpha\,,
\end{equation}
and taking the limit $\mu\rightarrow 0$, one arrives at the AdS$_5$ vacuum  \cite{Oz:1999qd},
\begin{equation}
\begin{split}
&d s_{11}^2 = 2^{-2/3}\,q_{\mathrm{M}5_1}^{2/3}q_{\mathrm{M}5_2}^{-2/3}\,c^{2/3}\left[ ds^2_{\text{AdS}_5}+ds^2_{\mathcal M_6}  \right]\,,\\
&ds^2_{\mathcal M_6}=d\alpha^2+s^2d\phi^2+4q_{\mathrm{M}5_1}^{-1}q_{\mathrm{M}5_2}^2c^{-2}(d\psi^2+d\chi^2)+4^{-1}c^2ds^2_{\tilde S^2}\,,\\
&G_{(4)}=-2^{-1}q_{\mathrm{M}5_1}q_{\mathrm{M}5_2}^{-1}sc^3d\alpha \wedge d\phi\wedge \text{vol}_{ \tilde S^2}-q_{\mathrm{M}5_2} d\psi\wedge d\chi \wedge \text{vol}_{ \tilde S^2}\,,
\end{split}
\end{equation}
with $ds^2_{\text{AdS}_5}=q_{\mathrm{M}5_1}^{-1}\,\mu^2ds^2_{\mathbb{R}^{1,3}}+\frac{d\mu^2}{\mu^2}$ and $s=\sin\alpha,\,c=\cos\alpha$.
The parameters of the solution can be related to those of the Type IIA vacuum \eqref{brane_metric_D4NS5_nearhorizon} through the identifications $q_{\mathrm{M}5_1}=q_{\mathrm{D}4}$ and $q_{\mathrm{M}5_2}=q_{\mathrm{NS}5}$. 
This solution is the uplift to 11d of the AdS$_5$ vacuum discussed in subsection \ref{AdS5semilocalD4NS5}, and it is therefore holographically dual, in M-theory, to the $\mathbb{Z}_n$ orbifold of 4d $\ma N=4$ SYM with $n=q_{\mathrm{M}5_2}/\sqrt{q_{\mathrm{M}5_1}}$.


Let us now introduce defects within the previous AdS$_5$ vacuum. The F1-D2-D4$'$-NS5$'$ defect branes considered in 
subsection \ref{branesystemIIA} become a M2-M5-M2$'$-M5$'$ intersection\footnote{The near horizon limit of the M2-M5-M2$'$-M5$'$ intersection was originally discussed in \cite{Boonstra:1997dy,Boonstra:1998yu} as an example of a standard brane intersection in M-theory reproducing $\mrm{AdS}_2\times S^2 \times \mathbb{R}^7$ close to the horizon.} completely localised within the worldvolume of the M5$_1$-M5$_2$ system. 
We consider the following eleven-dimensional metric, associated to the brane set-up depicted in Table \ref{Table:M5M5systemDefect}:
\begin{table}[http!]
\renewcommand{\arraystretch}{1}
\begin{center}
\scalebox{1}[1]{
\begin{tabular}{c c cc  c|| c c c  c |c c c}
 branes & $t$ & $x^1$ & $x^2$ & $x^3$ & $y$ & $z$ & $\psi$ & $\chi$ & $r$ & $\theta^1$ & $\theta^2$ \\
\hline \hline
$\mrm{M}5_1$ & $\times$ & $\times$ & $\times$ & $\times$ & $-$ & $-$ & $\times$ &$\times$ & $-$ & $-$ & $-$ \\
$\mrm{M}5_2$ & $\times$ & $\times$ & $\times$ & $\times$ & $\times$ & $\times$ & $-$ & $-$ & $-$ & $-$ & $-$ \\
$\mrm{M}2$ & $\times$ & $-$ & $-$ & $-$ & $\times$ & $-$ & $\times$ & $-$ & $-$ & $-$ & $-$ \\
$\mrm{M}5$ & $\times$ & $-$ & $-$ & $-$ & $-$ & $\times$ & $\times$ & $-$ & $\times$ & $\times$ & $\times$ \\
$\mrm{M}2'$ & $\times$ & $-$ & $-$ & $-$ & $-$ & $\times$ & $-$ & $\times$ & $-$ & $-$ & $-$ \\
$\mrm{M}5'$ & $\times$ & $-$ & $-$ & $-$ & $\times$ & $-$ & $-$ & $\times$ & $\times$ & $\times$ & $\times$ \\
\end{tabular}
}
\caption{BPS/8 intersection describing M2-M5-M2$'$-M5$'$ branes ending on a M5$_1$-M5$_2$ bound state. This system defined a $\ma N=4$ line defect within the $\ma N=2$ 4d CFT living in the  M5$_1$-M5$_2$ branes.}\label{Table:M5M5systemDefect}
\end{center}
\end{table}
\begin{equation}
\label{brane_metric_M5M5defect}
\begin{split}
d s_{11}^2 &= H_{\mathrm{M}5_1}^{-1/3}H_{\mathrm{M}5_2}^{-1/3}  \left[-H_{\mathrm{M}2}^{-2/3}H_{\mathrm{M}2'}^{-2/3}H_{\mathrm{M}5}^{-1/3}H_{\mathrm{M}5'}^{-1/3} dt^2+H_{\mathrm{M}2}^{1/3}H_{\mathrm{M}2'}^{1/3}H_{\mathrm{M}5}^{2/3}H_{\mathrm{M}5'}^{2/3}(d\rho^2+\rho^2ds_{S^2}^2) \right]\\
&+ H_{\mathrm{M}5_1}^{2/3}H_{\mathrm{M}5_2}^{-1/3}\left[H_{\mathrm{M}2}^{-2/3}H_{\mathrm{M}2'}^{1/3}H_{\mathrm{M}5}^{2/3}H_{\mathrm{M}5'}^{-1/3}dy^2+ H_{\mathrm{M}2}^{1/3}H_{\mathrm{M}2'}^{-2/3}H_{\mathrm{M}5}^{-1/3}H_{\mathrm{M}5'}^{2/3}dz^2  \right]\\
&+H_{\mathrm{M}5_1}^{-1/3}H_{\mathrm{M}5_2}^{2/3}\left[H_{\mathrm{M}2}^{-2/3}H_{\mathrm{M}2'}^{1/3}H_{\mathrm{M}5}^{-1/3}H_{\mathrm{M}5'}^{2/3}d\psi^2 +H_{\mathrm{M}2}^{1/3}H_{\mathrm{M}2'}^{-2/3}H_{\mathrm{M}5}^{2/3}H_{\mathrm{M}5'}^{-1/3} d\chi^2   \right]\\
&+H_{\mathrm{M}5_1}^{2/3}H_{\mathrm{M}5_2}^{2/3}H_{\mathrm{M}2}^{1/3}H_{\mathrm{M}2'}^{1/3}H_{\mathrm{M}5}^{-1/3}H_{\mathrm{M}5'}^{-1/3} \bigl(dr^2+r^2ds^2_{\tilde S^2}\bigr).
\end{split}
\end{equation}
In order to write the fluxes we take the M2-M5-M2$'$-M5$'$ branes fully localised within the worldvolume of the M5$_1$-M5$_2$ bound state. This implies that the warp functions $H_{\mathrm{M}2}$, $H_{\mathrm{M}2'}$, $H_{\mathrm{M}5}$ and $H_{\mathrm{M}5'}$ are only functions of $\rho$. We also take $H_{\mathrm{M}5_1}=H_{\mathrm{M}5_1}(y,z,r)$ and $H_{\mathrm{M}5_2}=H_{\mathrm{M}5_2}(r)$, for the M5$_1$-M5$_2$ bound state. The 4-form flux then takes the form,
\begin{equation}
\begin{split}\label{fluxes_M5M5defect}
G_{(4)}&=\partial_\rho H_{\mathrm{M}2}^{-1}dt\wedge d\rho\wedge dy \wedge d\psi+ \partial_\rho H_{\mathrm{M}5'}\,\rho^2\, \text{vol}_{  S^2}\wedge dz\wedge d\psi-\partial_\rho H_{\mathrm{M}2'}^{-1}dt\wedge d\rho\wedge dz\wedge d\chi\\
&+ \partial_\rho H_{\mathrm{M}5}\,\rho^2\, \text{vol}_{  S^2}\wedge dy\wedge d\chi+ \partial_r H_{\mathrm{M}5_2}\,r^2\,d\psi\wedge d\chi \wedge\text{vol}_{ \tilde S^2}+\partial_r H_{\mathrm{M}5_1}\,r^2\,  dy\wedge dz \wedge \text{vol}_{ \tilde S^2}\\
&+H_{\mathrm{M}2}H_{\mathrm{M}5}^{-1}H_{\mathrm{M}5_2}\partial_y H_{\mathrm{M}5_1}\,r^2\,  dz\wedge d r \wedge \text{vol}_{ \tilde S^2}
-H_{\mathrm{M}2'}H_{\mathrm{M}5'}^{-1}H_{\mathrm{M}5_2}\partial_z H_{\mathrm{M}5_1}\,r^2\,  dy\wedge d r \wedge \text{vol}_{ \tilde S^2}\,.
\end{split}
\end{equation}
We can now derive the equations of motion and Bianchi identities, observing that they split up into two groups. The harmonic functions associated to the M2-M5-M2$'$-M5$'$ defect branes turn out to be harmonic on $\mathbb{R}^2_\rho$,  and satisfying $H_{\mathrm{M}5}=H_{\mathrm{M}2}$, $H_{\mathrm{M}5'}=H_{\mathrm{M}2'}$ .
If we then consider the particular solution $H_{\mathrm{M}2}=1+\frac{q_{\mathrm{M}2}}{\rho}$ and $H_{\mathrm{M}2'}=1+\frac{q_{\mathrm{M}2'}}{\rho}$ and take the $\rho \rightarrow 0$ limit, we obtain the following class of solutions
\begin{equation}
\label{brane_metric_M5M5defect_nh}
\begin{split}
d s_{11}^2 &= q_{\mathrm{M}2}q_{\mathrm{M}2'}H_{\mathrm{M}5_1}^{-1/3}H_{\mathrm{M}5_2}^{-1/3}  \left(ds^2_{\text{AdS}_2}+ds^2_{S^2} \right)+ H_{\mathrm{M}5_1}^{2/3}H_{\mathrm{M}5_2}^{-1/3}\left(dy^2+dz^2  \right)\\
&+q_{\mathrm{M}2}q_{\mathrm{M}2'}^{-1}H_{\mathrm{M}5_1}^{-1/3}H_{\mathrm{M}5_2}^{2/3}\left(d\psi^2+d\chi^2\right)+H_{\mathrm{M}5_1}^{2/3}H_{\mathrm{M}5_2}^{2/3} \bigl(dr^2+r^2ds^2_{\tilde S^2}\bigr)\,,\\
G_{(4)}&=q_{\mathrm{M}2}\text{vol}_{\text{AdS}_2}\wedge dy \wedge d\psi-q_{\mathrm{M}2}\text{vol}_{\text{AdS}_2}\wedge dz\wedge d\chi- q_{\mathrm{M}2}\, \text{vol}_{  S^2}\wedge dz\wedge d\psi\\
&- q_{\mathrm{M}2}\, \text{vol}_{  S^2}\wedge dy\wedge d\chi+q_{\mathrm{M}2}q_{\mathrm{M}2'}^{-1} \partial_r H_{\mathrm{M}5_2}\,r^2\,d\psi\wedge d\chi \wedge\text{vol}_{ \tilde S^2}+\partial_r H_{\mathrm{M}5_1}\,r^2\,  dy\wedge dz \wedge \text{vol}_{ \tilde S^2}\\
&+H_{\mathrm{M}5_2}\partial_y H_{\mathrm{M}5_1}\,r^2\,  dz\wedge d r \wedge \text{vol}_{ \tilde S^2}
-H_{\mathrm{M}5_2}\partial_z H_{\mathrm{M}5_1}\,r^2\,  dy\wedge d r \wedge \text{vol}_{ \tilde S^2}\,,
\end{split}
\end{equation}
with $H_{\mathrm{M}5_1}$, $H_{\mathrm{M}5_2}$ satisfying the equations
\begin{equation}\label{EOM_M51M52}
 \nabla^2_{\mathbb{R}^3_r} H_{\mathrm{M}5_1}+H_{\mathrm{M}5_2}\nabla^2_{\mathbb{R}^2_{(y,z)}} H_{\mathrm{M}5_1}=0\qquad \text{and} \qquad  \nabla^2_{\mathbb{R}^3_r} H_{\mathrm{M}5_2}=0\,.
\end{equation}
Therefore, the $\rho \rightarrow 0$ limit of the background defined by \eqref{brane_metric_M5M5defect} gives rise to a new class of $\ma N=4$ solutions described by $\mrm{AdS}_2\times S^2$ geometries curving the worldvolume of the M5$_1$-M5$_2$ brane system.

In order to provide a defect interpretation to the solutions within this class  we choose the semi-localised solution for the M5$_1$-M5$_2$ branes given by \eqref{semilocalizedM51M52}, and make the change of coordinates \eqref{changeMtheory}. Taking the limit $\mu \rightarrow 0$ we obtain,
\begin{equation}
\begin{split}
&d s_{11}^2 = 2^{-2/3}\,q_{\mathrm{M}5_1}^{2/3}q_{\mathrm{M}5_2}^{-2/3}\,c^{2/3}\biggl[ \overbrace{q_{\mathrm{M}5_1}^{-1}\,q_{\mathrm{M}2}\,q_{\mathrm{M}2'}\mu^2\left(ds^2_{\text{AdS}_2}+ds^2_{S^2}\right)+\frac{d\mu^2}{\mu^2} }^{\text{locally}\,\,\, \text{AdS}_5\,\,\, \text{geometry}}+ds^2_{\mathcal M_6}\biggr]\,,\\
&ds^2_{\mathcal M_6}=d\alpha^2+s^2d\phi^2+4q_{\mathrm{M}5_1}^{-1}q_{\mathrm{M}5_2}^2q_{\mathrm{M}2}q_{\mathrm{M}2'}^{-1}c^{-2}(d\psi^2+d\chi^2)+4^{-1}c^2ds^2_{\tilde S^2}\,,\\
&G_{(4)}=q_{\mathrm{M}2}\left(\tilde c \text{vol}_{\text{AdS}_2}-\tilde s \text{vol}_{  S^2} \right)\wedge \left[\left(sd\mu+\mu c d\alpha \right)\wedge d\psi -\mu s d\phi \wedge d\chi \right]\\
&-q_{\mathrm{M}2}\left(\tilde s \text{vol}_{\text{AdS}_2}+\tilde c \text{vol}_{  S^2} \right)\wedge \left[\left(sd\mu+\mu c d\alpha \right)\wedge d\chi +\mu s d\phi \wedge d\psi \right]\\
&-\left ( 2^{-1}q_{\mathrm{M}5_1}q_{\mathrm{M}5_2}^{-1}sc^3d\alpha \wedge d\phi+q_{\mathrm{M}2}q_{\mathrm{M}2'}^{-1}q_{\mathrm{M}5_2} d\psi\wedge d\chi  \right) \wedge \text{vol}_{ \tilde S^2}\,,
\end{split}
\end{equation}
where $s=\sin\alpha$, $c=\cos \alpha$ and $\tilde s=\sin\phi$, $\tilde c=\cos \phi$.
Our analysis in this Appendix shows that if we choose the semi-localised solution \eqref{semilocalizedM51M52} and the  system of coordinates \eqref{changeMtheory}, it is possible to work out a limit in which the 11d background asymptotes locally to the AdS$_5$ geometry holographically dual, in M-theory, to a $\mathbb{Z}_n$ orbifold of $\ma N=4$ SYM. The M2-M5-M2$'$-M5$'$ branes  find in this way a line defect interpretation within this 4d CFT. This is no other but the strong coupling realisation of the Type IIA defect interpretation discussed in section 2.

\end{document}